\documentclass[5p]{elsarticle}
 
\usepackage{graphicx}  % figures
\usepackage{dcolumn}   % tables
\usepackage{multirow}
\usepackage{amssymb}   % math
\usepackage{amsmath}   % math
\usepackage{url}
\usepackage{comment}
\usepackage{hyperref} 
\hypersetup{breaklinks = true, colorlinks = true, citecolor = blue, linkcolor = blue, urlcolor = blue}

\def\bea{\begin{eqnarray}}
\def\eea{\end{eqnarray}}
\def\bean{\begin{equation*}}
\def\eean{\end{equation*}}

\begin{document}

\title{Neutron Spin Structure from  \texorpdfstring{e-$^3$He}{3He} Scattering with Double Spectator Tagging at the Electron-Ion Collider} 

\author[MIT,JLAB1,myfootnote,myfootnote1]{\texorpdfstring{I.~Fri\v{s}\v{c}i\'{c}}{I. Friscic}}

\author[MIT,JLAB1,myfootnote]{D.~Nguyen}

\author[MIT,JLAB1]{J.R.~Pybus}
\author[BNL]{A.~Jentsch}
\author[MIT]{E.P.~Segarra}
\author[PDS]{M.D.~Baker}
\author[MIT]{O.~Hen}
\author[JLAB1]{D.W.~Higinbotham}
\author[MIT]{R.~Milner}
\author[JLAB1]{A.S.~Tadepalli} 
\author[BNL]{Z.~Tu}
\author[JLAB1,LBNL]{J.~Rittenhouse~West}

\address[MIT]{Laboratory for Nuclear Science, Massachusetts Institute of Technology, Cambridge, MA 02139, USA}
\address[JLAB1]{Thomas Jefferson National Accelerator Facility, Newport News, Virginia 23606, USA}
\address[BNL]{Department of Physics, Brookhaven National Laboratory, Upton, NY 11973, USA}
\address[PDS]{Mark D. Baker Physics and Detector Simulations LLC, Miller Place, NY 11764, USA}
\address[LBNL]{Nuclear Science Division, Lawrence Berkeley National Laboratory, Berkeley, CA 94720, USA}

\fntext[myfootnote]{These authors contributed equally to this work.}
\fntext[myfootnote1]{Present address: Department of Physics, PMF, University of Zagreb, Croatia.}

\begin{abstract}
The spin structure function of the neutron is traditionally determined by measuring the spin asymmetry of inclusive electron deep inelastic scattering (DIS) off polarized $^3$He nuclei.
In such experiments, nuclear corrections are significant and must be treated carefully in the interpretation of experimental data. Here we study the feasibility of suppressing model dependencies by tagging both spectator protons in the process of DIS off neutrons in $^3$He at the forthcoming Electron-Ion Collider (EIC). This allows for a reconstruction of the momentum of the struck neutron to ensure it was nearly at rest in the initial state, thereby reducing sensitivity to nuclear corrections and suppressing contributions from electron DIS off protons in $^3$He. Using realistic accelerator and detector configurations, we demonstrate that the EIC can probe the neutron spin structure from $x_B$ of 0.003 to 0.651.
We find that the double spectator tagging method results in reduced uncertainties by a factor of $2$ on the extracted neutron spin asymmetries over all kinematics and by a factor of $10$ in the low-$x_B$ region, thereby providing valuable insight into the spin and flavor structure of nucleons.
\end{abstract}

\maketitle
%===================Introduction ===============================%
\section{Introduction}
The decomposition of nucleon spin in terms of quark spin, quark orbital angular momentum and total gluon angular momentum is a fundamental challenge in hadronic physics~\cite{Kuhn:2008sy,Leader:2013jra}. Experiments have found that quarks carry only around 30\% of the total spin of the nucleon \cite{1988PhLB..206..364A,RevModPhys.85.655} and future experiments are aimed at pinning down the contribution of the Orbital Angular Momentum (OAM) of the quarks in the valence region and gluon angular momentum \cite{Ji:2020ena}. Previous results hint at quark OAM contributions to the neutron spin~\cite{E99117-PRL}. Precise data is essential to test nucleon spin theories, in particular for the valence quark spin contributions where the contributions to spin from sea quark and anti-quark pairs as well as gluons are expected to be small \cite{Freese:2020mcx,Deur:2018roz}.

While stable particles such as hydrogen are readily available for studies of the proton structure, it is extremely challenging to make even low luminosity neutron beams ~\cite{COHERENT:2015mry}.  Instead, physicists have made use of deuterium and $^{3}$He targets as effective neutron targets in order to extract unpolarized nucleon structure functions $F_{1}$ and $F_{2}$~\cite{Bodek:1979rx,Arrington:2008zh}, polarized spin structure functions $g_{1}$ and $g_{2}$~\cite{Anthony:1993uf,Sulkosky:2019zmn}, as well as virtual photon asymmetry $A_1$~\cite{E99117-PRC}. In particular, $A_1$ is defined as $A_1 = (\sigma_{1/2} - \sigma_{3/2})/(\sigma_{1/2} + \sigma_{3/2})$ where subscript $1/2(3/2)$ is the projection of the total spin along the direction of the virtual photon momentum. Although polarized deuterium would allow for a single proton spectator tagged final state, it is not planned in the initial phase of Electron-Ion Collider (EIC) operations due to technical challenges related to the gyromagnetic ratio of the deuteron as compared to protons ($\rm \gamma_D/{\gamma_p} = 0.047$) and $^{3}$He ($\rm \gamma_D/{\gamma_{^3He}} = 0.081$) \cite{EICCDR}. On other hand, work is already underway on developing a polarized $^3$He source for the EIC \cite{Maxwell:2016knr} and it is within the scope of the initial project that polarized $^3$He beams will be able to be stored.

A novel experimental method for minimizing corrections due to off-shell neutron effects as well as deep inelastic scattering events from protons in $^3$He is to tag the recoiling spectator proton(s) from the target.   
This was done recently in the Jefferson Lab BONuS experiment~\cite{Baillie:2011za,Tkachenko:2014byy} with deuterium in which neutron deep inelastic scattering events were selected by requiring the detection of an accompanying low momentum proton.  
By ensuring the recoiling momentum of the proton is small, the experiment maximizes the probability that the electron has undergone deep inelastic scattering on a nearly free neutron~\cite{Cosyn:2014zfa}. 
This technique minimizes the model dependence of the extraction of the neutron information
and avoids many of the theoretical problems involved with extracting neutron information from inclusive scattering from a nuclear target~\cite{Cosyn:2019hem,Cosyn:2020kwu}.

\begin{figure}[htb]
\includegraphics[width=\linewidth]{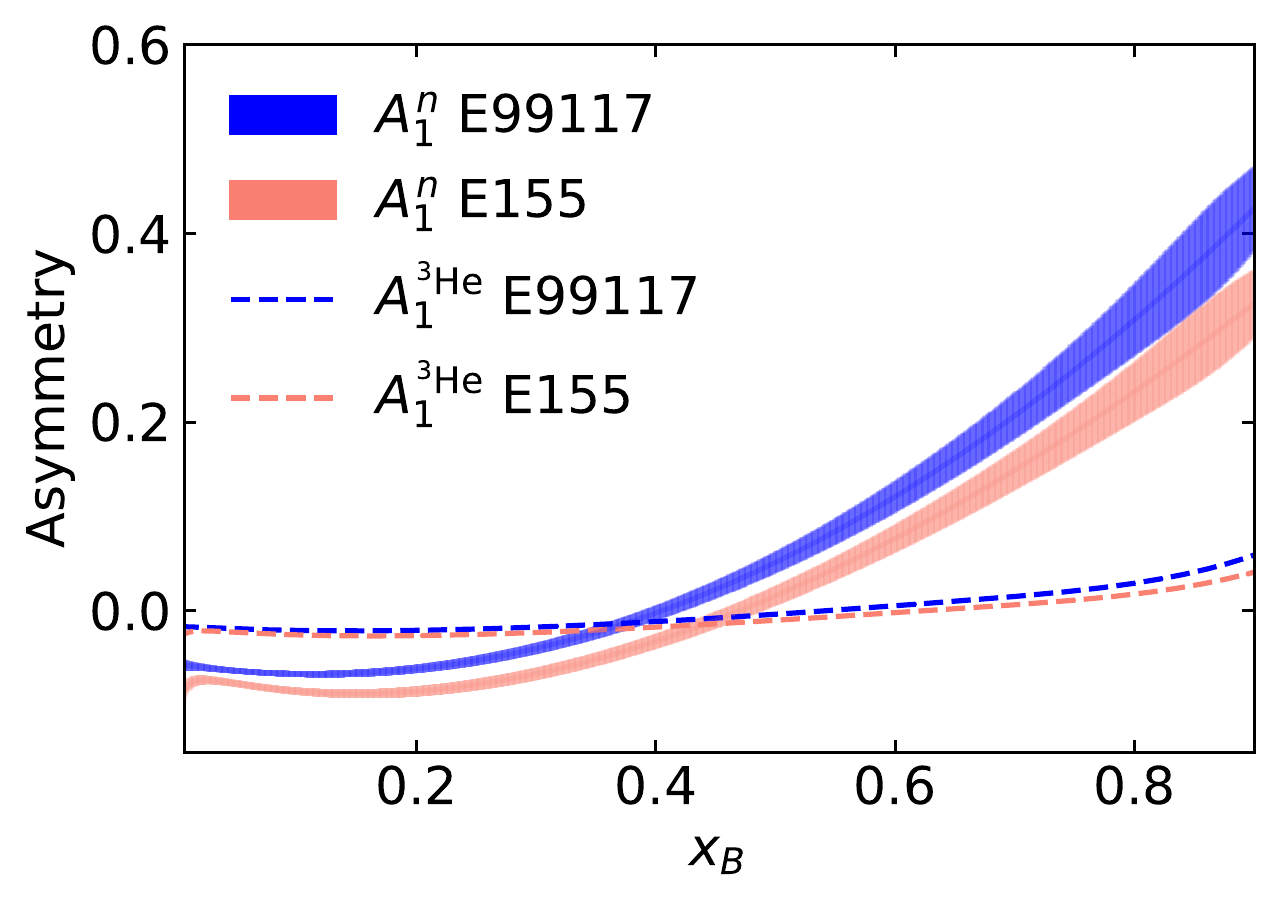}
\caption{The asymmetries $A_1^n$ (bands) and $A_{1}^{^3\text{He}}$ (dashed-line) as a function of $x_{B}$ at $Q^2 = 5$ (GeV$/$c)$^2$. The two $A_{1}^{^3\text{He}}$ curves were calculated using $A_{1}^{p}$ parameters in~\cite{E99117-PRC} and $A_1^n$ parameters in~\cite{E99117-PRC} [E99117] and~\cite{Anthony:2000fn} [E155] following a simple estimation based on formula \ref{a1he}. The bands show the extracted $A_1^n$ from $A_{1}^{^3\text{He}}$ as described in section \ref{a1n-extraction}.}
\label{fig:A1n-A1He3}
\end{figure}

Unfortunately, this same technique has never been done with
fixed target polarized $^{3}$He targets, where the low momentum recoil protons would need to pass through the glass of the target cell walls in order to be detected.
Thus, to date, the neutron structure functions are extracted from $^3$He measurements and require large model dependent corrections from the measured cross sections and asymmetries to extract the quantities of interest~\cite{Sulkosky:2019zmn, Flay:2016wie}.
In Fig.~\ref{fig:A1n-A1He3} we show the large difference between the measured quantities and the extracted neutron information.

While our knowledge of the three-body system and nuclear corrections is good, it is clearly not perfect as has been shown experimentally~\cite{Mihovilovic:2018fux,Mihovilovic:2014gdi,Long:2019iig, Cruz-Torres:2020uke}. Traditionally, one measures the DIS off the polarized $^{3}$He to determine the virtual photon asymmetry $A_{1}^{^3\text{He}}$, and then deduces the neutron asymmetry $A_{1}^{n}$.
In this procedure the model dependent proton polarization used as an input in extraction dominates the total systematic uncertainty \cite{E99117-PRC}. Therefore, while accurate measurements of $A_{1}^p$ exist for the proton, more precise data on effective polarized neutron targets are necessary.

With high energy electron and $^3$He beams at the EIC, it is possible to detect recoiling spectator protons with low momentum in the rest frame of $^{3}$He. These recoil protons, which have a high momentum in the lab frame, can be detected in the far forward region of the EIC. This unique tagging capability will ensure that the electron interacted with a nearly free initial-state neutron and allows an extraction of neutron spin structure with less model-dependence and more precision than inclusive techniques.

\section{DIS Kinematics with Double Spectator Tagging}
High energy electron beams have made it possible to access DIS kinematics. This region corresponds to scattering well beyond the nucleon resonances~\cite{Friedman:1972sy} and is experimentally accessed by going to large four-momentum squared transfer $Q^2\equiv -q^2\geq 2$ (GeV$/$c)$^2$ and large hadronic system invariant mass $W^{2}\geq 4$ (GeV$/$c)$^2$.

A DIS electron-nucleon event is defined by the initial four-momentum of the electron and nucleon $k=(E_e, \vec{k}_e)$, $p = (E_p, \vec{p})$, respectively, the final four-momentum of the electron  $k'= (E^{'}_e,\vec{k}^{'}_e)$ and the four-momentum transfer $q=k-k'$. 
In the case of electron-nucleus scattering, $p$ defines the four-momentum of an on-mass-shell nucleon at rest within the nucleus.
We define the following quantities for the event:
\begin{itemize}
  \item the Bjorken variable $x_B=\frac{Q^2}{2(q\cdot p)}$, where $Q^2= -q^2$ is minus the four-momentum transfer squared;
  \item the fractional energy transfer $y=\frac{q\cdot p}{k\cdot p}$;
  \item the final hadronic system squared invariant mass $W^2=(q+p)^2$; 
  %, where the target is assumed to be a nucleon at rest in the nuclear rest frame;
  \item the energy of the beam electron $E_{e}$;
  \item the energy $E_e'$, opening angle $\theta_e'$, and azimuthal angle $\phi_e'$ of the scattered electron.
\end{itemize}

\begin{figure}[htb]
    \centering
    \includegraphics[scale=0.3]{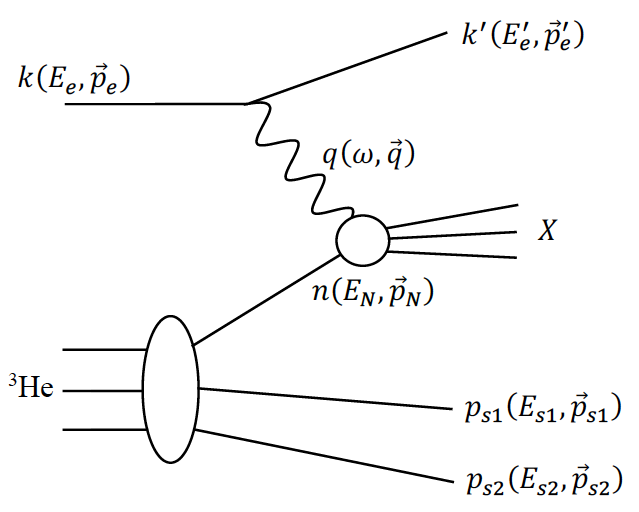}
   
    \caption{A diagram of Deep Inelastic $e$+$^3$He scattering with double spectator tagging. The channel shown here is electron scattering off a neutron in $^3$He; the two spectator nucleons are the protons in the process $^3$He$(e,e'p_{s1}p_{s2})X$.}
    \label{fig:diagram}
\end{figure}

For the nuclear system, we define the initial four-momentum of the nucleus $p_A$ and the initial four-momentum of the nucleons $p_i$.
Here $i=1,s1,s2$, where nucleon $1$ is the struck nucleon and nucleons $s1$ and $s2$ are the spectator nucleons, as shown in Fig~\ref{fig:diagram}. For each nucleon we also define, in the nuclear rest frame:
\begin{itemize}
    \item the momentum perpendicular to the momentum transfer $\bf p^\perp_i$;
    \item the ``plus/minus" component of momentum $p_i^\pm=E_i\pm p_i^z$, where the $z$-component is defined parallel to the momentum transfer;
    \item the light-front momentum fraction $\alpha_i=\frac{A}{m_A}p_i^-$ where $A$ is the atomic mass number and $m_A$ is the target mass. 
\end{itemize}
%These momenta satisfy $\sum_i {\bf p_i^\perp}=\bf 0$ and $\sum_i p_i^-=m_A$, in order to preserve conservation of transverse and minus momentum. 
Finally, we define the virtuality of the struck nucleon $v=m_N^2 - (p_A - p_{s1} - p_{s2})^2$.

\begin{figure}[htb]
\includegraphics[width=\linewidth]{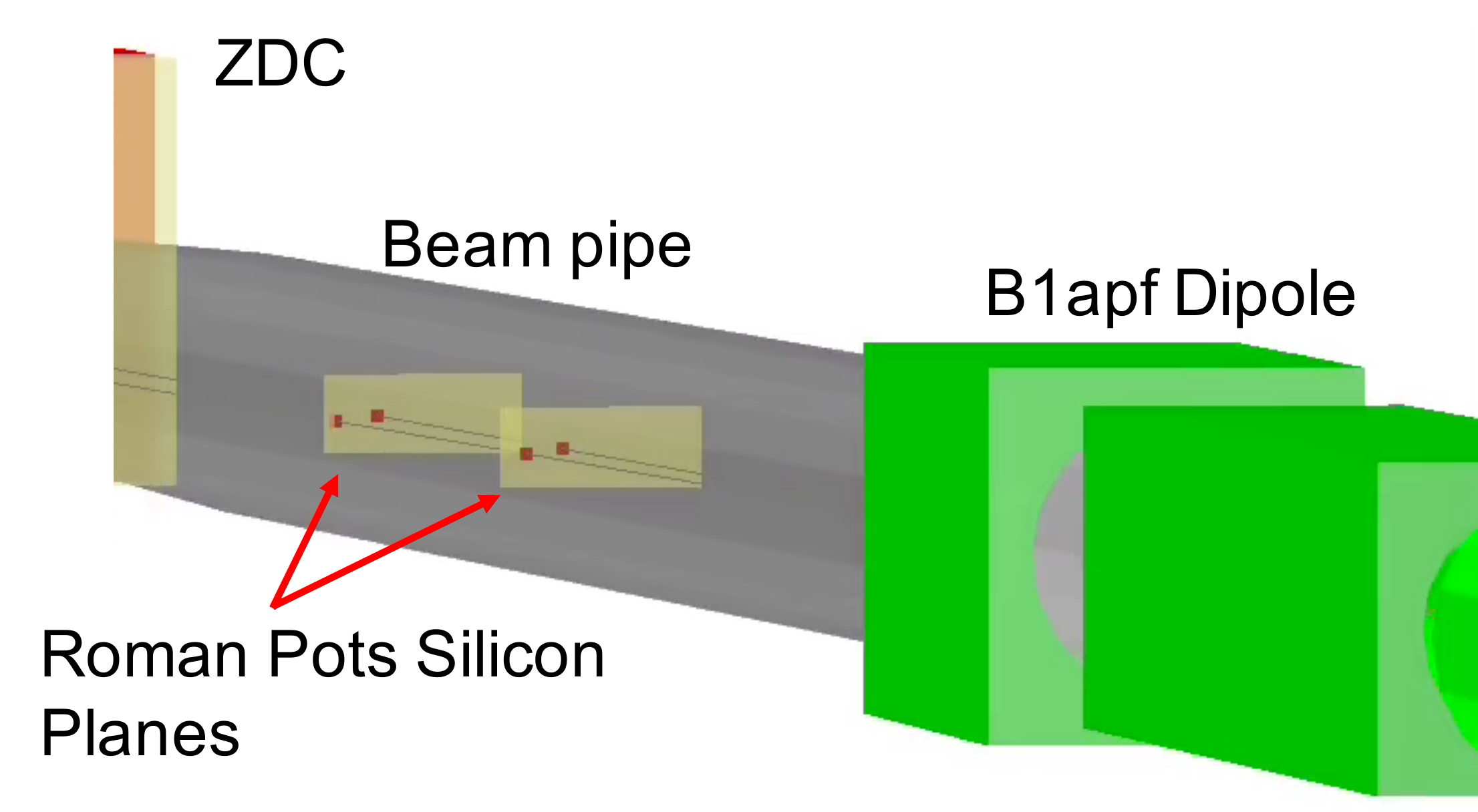}
  \caption{Zoomed-in image of a 3BBU-MF e+$^{3}\rm{He}$ with both protons tagged in the Roman Pot silicon detector planes. The beam pipe was made semi-transparent in order to see the event. The off-momentum detectors are omitted from the figure to make the event easier to view. }
  \label{fig:figure_IR_RP_zoom_in}
\end{figure}

\section{Event Generator and Fermi Motion Correction}
%===I just added this to add the reference===//
DIS events were generated using the CLASDIS generator, a CLAS version of PEPSI~\cite{Mankiewicz:1991dp} generator, which is based on LEPTO version 6.5 and JETSET version 7.410. The generator is also capable of generating semi-inclusive DIS (SIDIS) events. The CLASDIS generator is intended to be used for the fixed target event generation (electron beam goes along $+z$ direction); in order to use it for the EIC kinematics (ion beam goes along $+z$ direction and the electron beam goes into opposite direction), the selected EIC kinematics in terms of electron and ion beam energies need to be boosted in the fixed target frame, in which the ion beam is at rest and the energy of the electron beam is Lorentz-boosted. Events are then generated
in the fixed target frame using the CLASDIS generator and boosted back into the collider frame of the EIC.

CLASDIS can generate (SI)DIS events for polarized and unpolarized targets, but is currently only capable of generating DIS events from free protons and free neutrons.
CLASDIS approximates $^{3}$He electron scattering events by combining the scattering from three free nucleons (two protons, one neutron).
This approximation does not include the effects of the Fermi motion of the nucleons inside the nucleus on generated DIS events, nor are the kinematics of the spectators (nucleons or nuclei) considered. 
A separate procedure was therefore developed to account for each of these effects.

Nuclear corrections to DIS were modeled within three breakup regimes depending on the momentum of the struck nucleon and the excitation of the residual system. 
The three breakup regimes were classified as 2-body breakup (2BBU), mean-field 3-body breakup (3BBU-MF), and short-range correlation breakup (SRC). 

In the 2BBU regime momentum of the leading nucleon is given by the Ciofi-Kaptari spectral function~\cite{CiofidegliAtti:2005qt}, which constrains the momentum of the residual nucleus. For the 3BBU-MF regime, the momentum of the leading nucleon and the energy of the spectator system were taken from the Ciofi-Kaptari spectral function~\cite{CiofidegliAtti:2005qt}, which is truncated at a leading nucleon momentum of 240 MeV/c to provide separation from the SRC-dominated regime and ensure the predicted SRC pair fractions. The magnitude of the relative momentum of the spectator system was determined from the system's energy, and the direction was sampled assuming an isotropic distribution. In the case of the SRC regime, the relative momentum and center-of-mass momentum of the leading pair were determined using the light-front Generated Contact Formalism (GCF)~\cite{Pybus:2020itv}, which constrained the momentum of the residual nucleon.

The relative strengths of the 2BBU regime~\cite{CiofidegliAtti:2005qt} and SRC regime~\cite{Cruz-Torres:2019fum} were used to constrain the population of the 3BBU-MF regime, as listed in table~\ref{tab:3He_strength}.

\begin{table}[htb]
\centering
\caption{The relative strength of breakup regimes}
\label{tab:3He_strength}
\begin{tabular}{ccc}
\hline
Regime  & Proton Fraction & Neutron Fraction \\ 
\hline
2BBU    & $67\%$ & \textemdash \\
3BBU-MF & $27\%$ & $90\%$ \\
SRC     & $6\%$  & $10\%$ \\     
\hline
\end{tabular}
\end{table}

In order to add nuclear effects to each CLASDIS generated events, calculations were conducted on the light-front, meaning  that  the  light-front fraction and transverse momentum for each nucleon was required, rather than the standard 3-momentum. The GFC calculations already included this~\cite{Pybus:2020itv}, but for the Ciofi-Kaptari spectral  function  it  was  necessary  to  move  the  initially sampled  kinematics  to  the  light-front.  The  Frankfurt-Strikman  formalism~\cite{Frankfurt:1981mk}, developed for  the  deuteron on the light-front, was generalized as follows:
for each particle, the momentum was determined, and the energy was selected to place the particle on-shell, resulting in temporary violation of energy conservation ($\sum_i E_i \neq m_{A}$). The system was then Lorentz boosted into a frame where the minus momentum ($p^-\equiv E-p_z$) was conserved ($\sum_i p^-_i = m_{A}$), leaving the plus momentum ($p^+\equiv E+p_z$) unconserved ($\sum_i p^+_i \neq m_{A}$). This provided satisfaction of the light-front baryon and momentum sum rules~\cite{Frankfurt:1981mk}. The effects of this boost were relatively small, resulting in nonconservation of the plus momentum on the order of 30 MeV$/$c. Light-front effects were more significant in the high-momentum SRC regime, where the GCF calculations were performed manifestly on the light-front.

Events taken from the CLASDIS generator were adjusted to account for nuclear motion. Considering only inclusive scattering, the free-nucleon events are described by a cross section
\begin{equation}
    \frac{d^3\sigma}{dx_Bdyd\phi_e'}\equiv \sigma_{free}(x_B,y,\phi_e')
\end{equation}
described in~\cite{Mankiewicz:1991dp}.
By altering the kinematics to replace the Bjorken scaling variable $x_B$ with the scaling variable $x'=x_B/\alpha_1$ which includes the smearing effects of nuclear motion, we change the effective cross section to one which performs a convolution between the free DIS cross section and the light-front nuclear spectral function:
\begin{equation}
\begin{split}
    \frac{d^9\sigma}{dx'dyd\phi_e'\frac{d\alpha_{s1}d^2{\bf p_{s1}^\perp}}{\alpha_{s1}}\frac{d\alpha_{s2}d^2{\bf p_{s2}^\perp}}{\alpha_{s2}}}&= \nonumber \\  \sigma_{free}&(x',y,\phi_e')S(\alpha_{s1},{\bf p_{s1}},\alpha_{s2},{\bf p_{s2}})
\end{split}
\end{equation}
where $S(\alpha_{s1},{\bf p_{s1}},\alpha_{s2},{\bf p_{s2}})$ is the light-front spectral function describing the structure of the nucleus, including all regimes previously described. After sampling the spectral function, modifying the CLASDIS events to account for the nuclear motion, and adding the spectator protons, we receive a sample of tagged DIS events described by the above cross section.

\section{Full Detector Simulations with EicRoot} \label{sec:fullSimSection}
%\section{Simulation in EicRoot framework}\label{sec:fullSimSection}

Realistic detector simulations have been carried out using the EIC reference detector far-forward (FF) configuration implemented in the EicRoot simulation framework, which makes use of the ROOT Virtual Monte Carlo structure and GEANT4~\cite{GEANT4}.

The FF detector subsystems, shown in Fig. \ref{fig:figure_IR_RP_zoom_in}, are optimized to try and cover as much of the acceptance at $\eta > 4.5$ as possible. Table \ref{tab:FFDetectors_acceptance} summarizes the geometric acceptances and ranges of typical absolute transverse momentum smearing for the four FF detector subsystems, each of them covering different regions of the FF acceptance and a  unique  area of the  phase  space. 

\begin{table}[thb]
%\fontsize{12}{15}\selectfont
\fontsize{9}{13}\selectfont
  \caption{\label{tab:FFDetectors_acceptance} Summary of 
  basic polar scattering angle acceptance and average $p_{T}$ smearing for each FF detector subsystem.  Details can be found in the EIC Yellow Report~\cite{AbdulKhalek:2021gbh}. }
\begin{center}
\begin{tabular}{lcccc}
\hline
  \textbf{Detectors} & $\theta$ Acc. & $p_{T}$ res. \\
   & [mrad] &  [MeV/c]  \\
\hline
 B0 tracker &  [5.5, 20.0]  & 15-30 \\
 Off-momentum &  [0.0, 5.0]   & 20-30  \\
 Roman Pots &  [0.0, 5.0]  & 20-30  \\
 ZDC &   [0.0, 4.5]  & 30-40 \\
\hline
 \end{tabular}
 \end{center}
 \end{table}

The first detector along the hadron beamline after the interaction point is the B0 detector, which is comprised of silicon tracking layers inside the B0pf dipole magnet. This subsystem is used for capturing charged particles at larger scattering angles ($\theta > 5.5$ mrad). 

The Roman Pots are situated $\sim27$ meters downstream from the interaction point. They consist of silicon sensors placed in Roman Pot vessels or RF shields which are injected into the beam line vacuum a few mm from the hadron beam. The Roman Pots detector subsystem is used for capturing protons with smaller scattering angles ($\theta < 5$ mrad), but with a momentum similar to that of the settings for the steering magnets, where the tagged proton momentum vs. beam magnet setting is referred to as the magnetic rigidity.

The Off-Momentum Detectors were designed to optimally tag charged particles with a rigidity $\sim\frac{1}{2}$ that of the beam, but otherwise have a similar acceptance and smearing contribution to that of the Roman Pots. They are placed outside the beam pipe in the same drift region as the Roman Pots detectors. This enables tagging of charged particles as they are bent out of the beam pipe by the last dipole magnet (B1apf) before the drift region. 

The Zero-Degree Calorimeter (ZDC) will contain both hadronic and electromagnetic calorimetry and is used for tagging neutrons from nuclear breakup, as well as photons from coherent nuclear scattering. 

Various effects impact the overall measured smearing in the momentum reconstruction. There are contributions from the detector assumptions which include the finite pixel size of the silicon detectors used in the B0 detector, Roman Pots detectors, and the Off-Momentum Detectors, as well as the assumptions used in the reconstruction (e.g. a linear transport matrix for the Roman Pots). There are also effects from the beam such as angular divergence and vertex smearing induced by the crab cavities used to compensate for the EIC crossing angle of $25$ mrad.
The details of the detector layout and assumptions can be found in the EIC Yellow Report~\cite{AbdulKhalek:2021gbh}, in a previous study of exclusive $J/\psi$ production~\cite{Tu:2020BeAGLE} and Conceptual Design Report~\cite{EICCDR}.

\section{Fast Simulations with Eic-smear}

Eic-smear is a Monte Carlo analysis package, which was developed by the BNL EIC task force. It is a collection of ROOT classes and routines for analysing Monte Carlo events and performing fast smearing to study the effects of detector resolution and it is optimized for processing of large amount of data. 
For our particular channel of interest the scattered electron is smeared with smearing functions for the central detector in the pseudo-rapidity range $-3.5 < \eta < 3.5$ and the two tagged protons in the far-forward region $\eta > 4.5$. The central detector consist of several different detector systems which cover their own range of pseudo-rapidity $\eta$ and corresponding resolution of three-momentum $p$ (tracking) and energy $E$ (EM calorimeters). For tagged protons only the three-momentum $p$ and transverse momentum $P_t$ are smeared. The smearing is summarized in table \ref{tab:SM}), which parameters correspond to the most updated version of the smearing matrix from November 21 2020 \cite{matrix02}, as defined by the Yellow Report Detector working group. 

\begin{table}[thb!]
\fontsize{9}{13}\selectfont
  \caption{\label{tab:SM} Summary of relevant particle smearing configuration performed in Eic-smear software package, as defined in Matrix 02 \cite{matrix02}, the newest (November 21 2020) detector matrix released by the EIC Yellow Report Detector working group.}
  \begin{center}
  \begin{tabular} {cc} \hline
  $\eta$ range & Tracking \\ \hline
  -3.5 --  -2.5 & $\sigma_p/p \sim 0.2\% \cdot p +5\% $ \\
 -2.5 --  -1.0   & $\sigma_p/p \sim 0.04\% \cdot p+ 2\%$ \\
 -1.0 -- 1.0  & $\sigma_p/p \sim 0.04\% \cdot p+ 1\%$ \\
  1.0 -- 2.5  & $\sigma_p/p \sim 0.04\% \cdot p+ 2\%$ \\
  2.5 -- 3.5  & $\sigma_p/p \sim 0.2\%  \cdot p+ 5\%$ \\ \hline
                   & EM Calorimeters \\ \hline
-3.5 -- -2.0  & $1\%/E +2.5\%/\sqrt{E} +1\%$ \\
-2.0 -- -1.0  & $2\%/E +8\%/\sqrt{E} +2\%$ \\
-1.0 -- 1.0  & $2\%/E +14\%/\sqrt{E} +3\%$ \\
1.0 – 3.5  &  $2\%/E +12\%/\sqrt{E} +2\%$ \\ \hline
%                   & Hadronic Calorimeters \\ \hline
%-3.5 -- -1.0  & stochastic $= 50\%$, const. $= 10\%$   \\
%-1.0 -- 1.0  & stochastic $= 100\%$, const. $= 10\%$   \\
%1.0 – 3.5  & stochastic $= 50\%$, const. $= 10\%$  \\ \hline
\multicolumn{2}{c}{Far-Forward region: protons } \\ \hline
\multicolumn{2}{c}{$\sigma_p/p= 5\%$ and  $\sigma_{P_t}/P_t = 3\%$} \\ \hline  
\end{tabular}
\end{center}
\end{table}

\section{Selected EIC Kinematics}

\begin{figure*}[ht!]
    \centering
    \includegraphics[width=0.49 \linewidth]{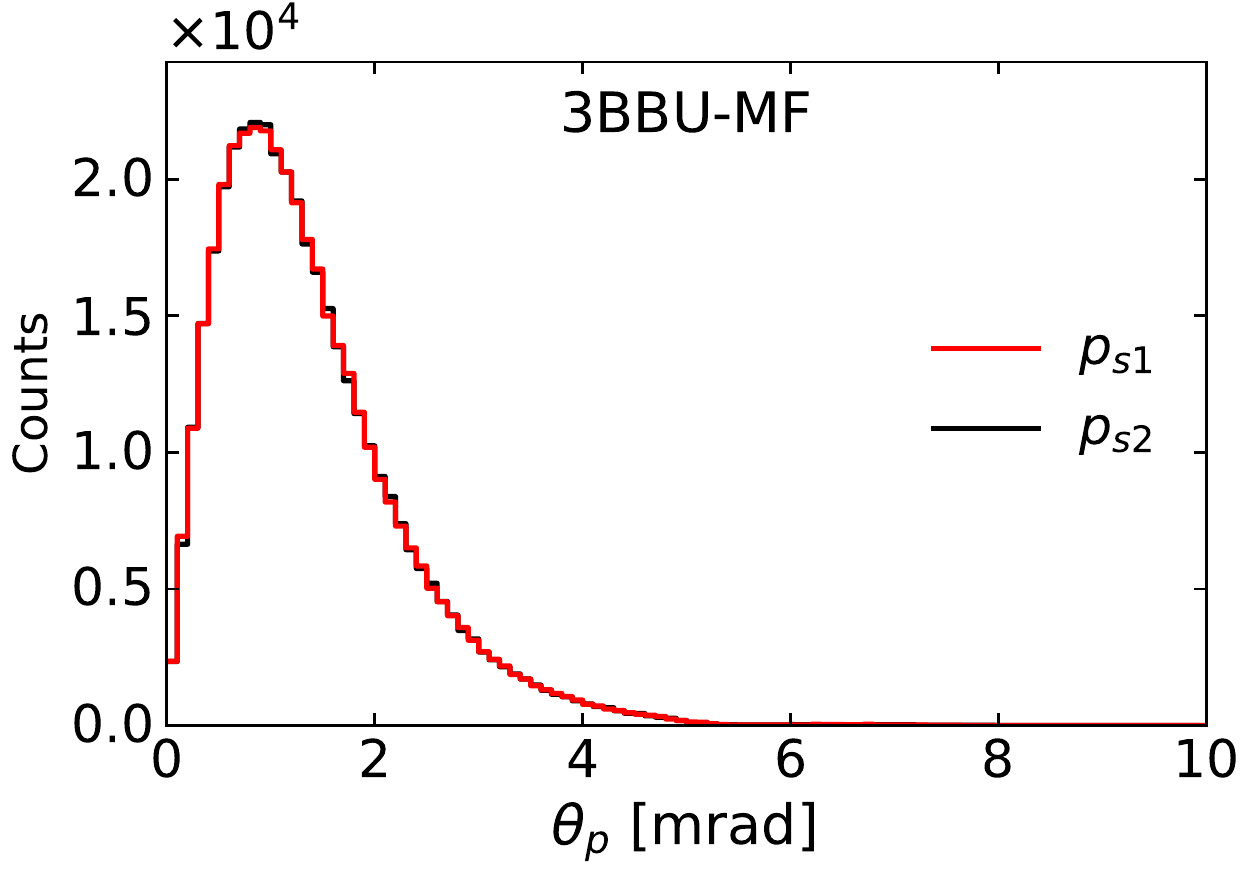}
    \includegraphics[width=0.49 \linewidth]{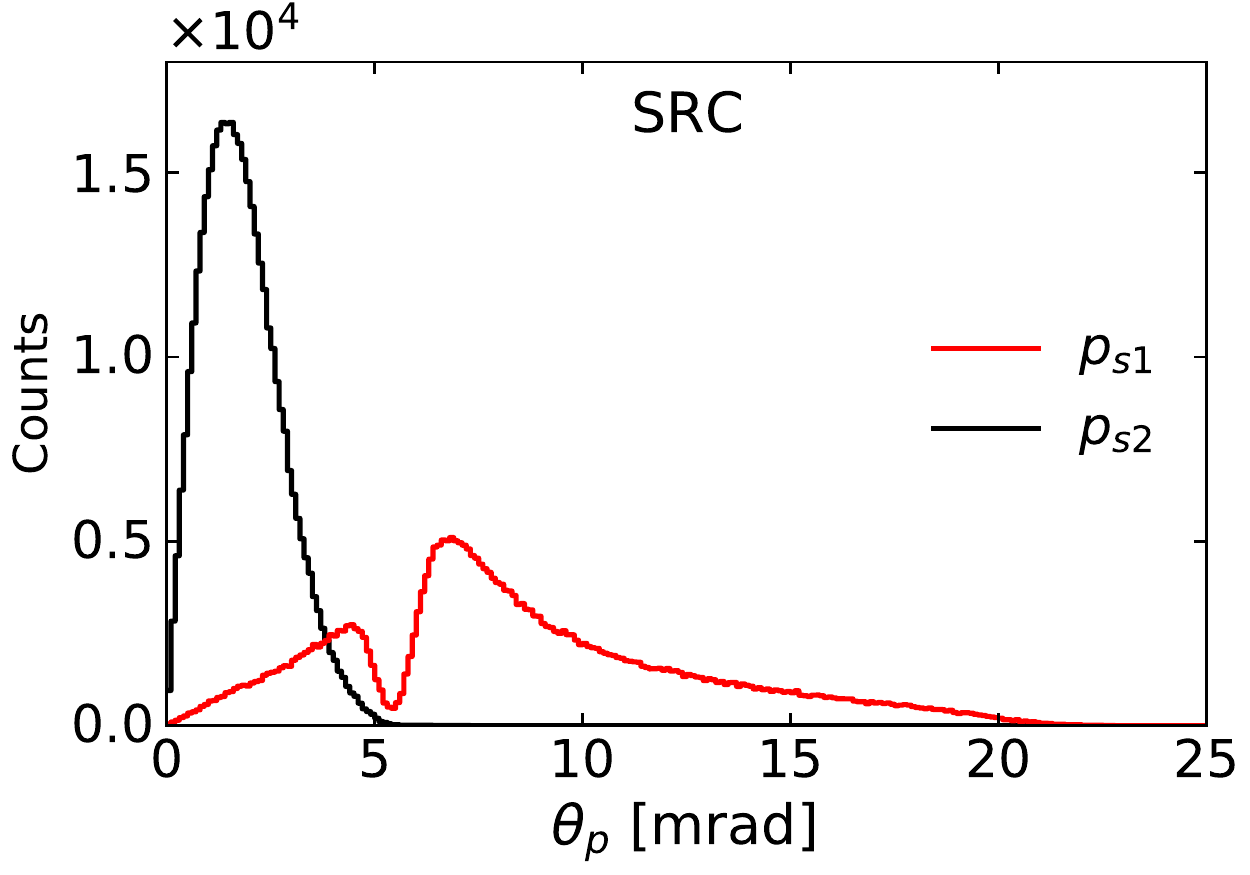}
    \caption{Distribution of the scattering angle $\theta_p$ of both spectator protons in case of the 3BBU-MF and SRC nuclear effects. The gap in the angular distribution for the SRC case is due to gap in the detector acceptance in the far-forward region, as it is summarized in table \ref{tab:FFDetectors_acceptance}. }
    \label{pp-41x5}
\end{figure*}

For the purpose of this paper, we have selected one of the two 
beam energies combinations from the Yellow Report~\cite{AbdulKhalek:2021gbh}. The chosen electron beam energy was 5 GeV and $^{3}$He beam had 41 GeV/nucleon (5x41). Since the CLASDIS generator assumes the fixed target kinematics, the electron beam was boosted into the frame in which the $^{3}$He is at rest.  In order to be in the DIS regime, events were generated with the following cuts: $Q^2 >$2 (GeV$/$c)$^2$, $W^2 >$ 4 (GeV$/$c)$^2$, $z>0.2$ and $0.05 <y <$ 0.95. Fermi corrections were applied to the generated events in the fixed target frame, which were then boosted into the collider frame. These events were then processed using the two different simulation frameworks EicRoot and Eic-smear. Figure \ref{pp-41x5} shows the angular distribution of both tagged protons scattering angle $\theta_p$, for the each of the two relevant nuclear effects, 3BBU-MF and SRC, separately. 

%\section{Physics Measurements}
\section{Simulation of the Double Spectator Tagging with FF Detectors}

\begin{figure}[ht!]
    \centering
    \includegraphics[width=\linewidth]{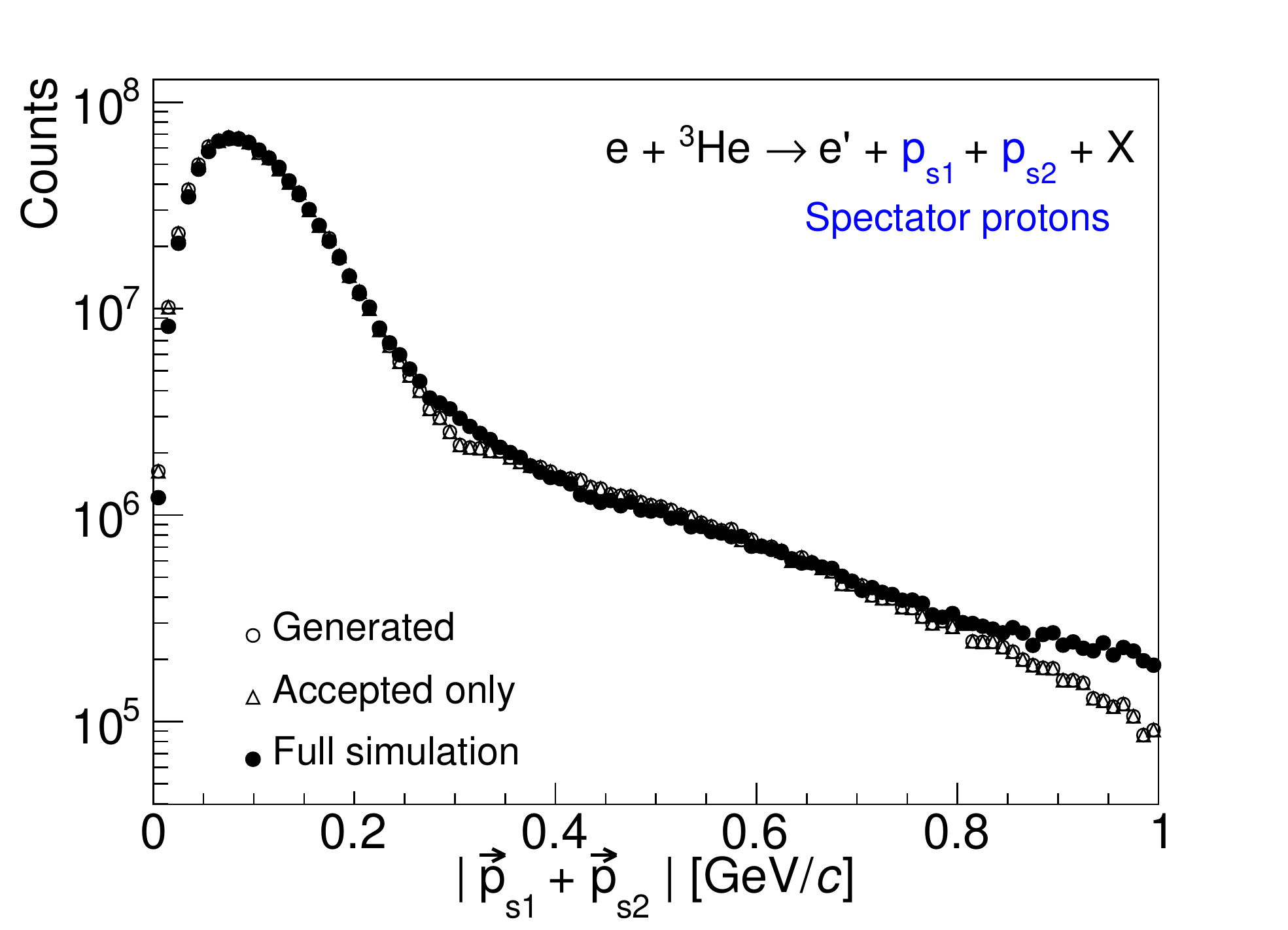}
    \caption{Distribution of the vector sum of two spectator protons' momentum, $p_{s1}$ and $p_{s2}$, in the fixed target frame for kinematic setting 5x41 GeV/n. The simulation is shown at the generated level, including acceptance correction, and including full simulation with smearing effects, using open circles, open triangles and the solid points, respectively. In general, for both 3BBU-MF and SRC protons included in this figure, the acceptance is $\sim$90\% or better except in the gap regions between detector subsystems.}
    \label{He3-41x5}
\end{figure}

The full simulation framework EicRoot was used to study the performance of detectors
which are intended to be used in the FF region for detection of two spectator protons. 200k DIS events were generated with the CLASDIS. SRC and 3BBU-MF regimes were added separately and in total 400k events were processed with EicRoot. Information provided by the EicRoot was used to reconstruct momenta of two spectator protons. Events from the SRC and 3BBU-MF regimes were weighted with their respective strengths (table \ref{tab:3He_strength}) and finally combined to obtain the simulaton of the spectator double tagging with FF detectors of the $^3$He$(e,e'p_{s1}p_{s2})X$ channel. The sum of momenta of two spectator protons in the fixed target frame is approximately equal to the initial momentum of the struck neutron in $^3$He, as presented in Fig.~\ref{He3-41x5}.

By constraining these spectator nucleons at small angle (along the ion beam direction) in the collider frame and their total momentum close to zero in fixed target frame, the DIS events from a ``free" neutron with minimal off-shell effects can be selected.
By tagging two low momentum protons after electron and polarized $^3$He beam scattering, one can minimize nuclear corrections and gain unique access to:
\begin{itemize} 
\item the asymmetry $A_{1}^{^3\text{He}}$ and $A_1^n$ as a function of $x_{B}$;
\item the neutron structure function $F_{2}^{^3\text{He}}$ and $F_{2}^n$ as a function of $x_{B}$.
\end{itemize}

\iffalse
\begin{figure}[ht!]
    \centering
    \includegraphics[width = \linewidth]{Figures/New_plots/smeared_bins.pdf}
    \caption{The truth and smeared $x_B$ distribution. Only electron momentum information was used to reconstruct smeared $x_B$}
    \label{smear-NM}
\end{figure}
\fi

%\section{Results}{\label{a1n-extraction}} 
\section{Inclusive DIS vs. Inclusive DIS with Double Spectator Tagging}{\label{a1n-extraction}} 
Experimentally, the virtual photon asymmetry $A_1$ can be extracted from the measured longitudinal electron asymmetry $A_{||}$ and transverse electron asymmetry $A_{\perp}$ where 
\begin{equation*}
    A_{||}= \frac{\sigma_{\downarrow\Uparrow} 
    - \sigma_{\uparrow\Uparrow}}
    {\sigma_{\downarrow\Uparrow}
   + \sigma_{\uparrow\Uparrow}}
     \quad \text{and} \quad
    A_{\perp} = \frac{\sigma_{\downarrow\Rightarrow} - \sigma_{\uparrow\Rightarrow}}{\sigma_{\downarrow\Rightarrow} + \sigma_{\uparrow\Rightarrow}}.
\end{equation*}

Considering electromagnetic interaction only,
$\sigma_{\downarrow\Uparrow}(\sigma_{\uparrow\Uparrow})$ is the cross section of the electron spin anti-parallel (parallel) to beam direction scatter off the longitudinally polarized target. $\sigma_{\downarrow\Rightarrow} (\sigma_{\uparrow\Rightarrow})$ is the cross section of the electron spin ani-parallel (parallel) scatter off the transversely polarized target. The relation between $A_1$, $A_{||}$ and $A_{\perp}$ is 
\begin{equation}
A_{1}=\frac{A_{\|}}{D(1+\eta \xi)}-\frac{\eta A_{\perp}}{d(1+\eta \xi)},
\label{A1_formula}
\end{equation}

where $D = {y(2-y)(2+\gamma^2y)}/(2(1+\gamma^2)y^2 + (4(1-y) - \gamma^2y^2)(1+R))$, $d= \sqrt{4(1-y)-\gamma^2y^2}D/(2-y)$, $\eta = \gamma(4(1-y) - \gamma^2y^2)/(2-y)/(2+\gamma^2y)$, $\xi = \gamma(2-y)/(2 + \gamma^2y)$, $\gamma = 4M^2x^2/Q^2$ \cite{Sato:2016tuz, E99117-PRC} and $R$ is the ratio of the longitudinal and transverse virtual photon absorption cross sections $\sigma_{L} / \sigma_{T}$ \cite{Feynman:1973xc}. The world fit parameters in Ref.\cite{E143:1998nvx} are used to calculate value of $R$. However, the analysis within this work focuses only on $A_1^n$.

The procedure applied in extracting $A_1^n$ from $A_1^{^3\text{He}}$ through inclusive $(e,e')$ measurement and directly measured $A_1^n$ using the double spectator tagging method are described in this section.
\begin{itemize}
\item The values of $A_1^n$ and $A_1^p$ are taken from the parameterization provided in~\cite{E99117-PRC}. The uncertainties and the correlation matrix associated with $A_1^n$ and $A_1^p$ parameterization have also been obtained from~\cite{E99117-PRC}. 

\item The structure functions $F_2^p$ and $F_2^\text{D}$ are taken from the world data fit NMC E155 \cite{Arneodo:1995cq}. The larger of the asymmetric uncertainties is chosen as the symmetric uncertainty for these structure functions.

\item Assuming no off-shell or nuclear-motion corrections, the value of $F_2^n$ is obtained using $F_2^n = F_2^\text{D} - F_2^p$. Similarly, $F_2^{^3\text{He}}$ is obtained by using $F_2^{3\text{He}} = F_2^\text{D} + F_2^p$. The uncertainties of $F_2^n$ and $F_2^{^3\text{He}}$ are propagated from the uncertainties of $F_2^\text{D}$ and $F_2^p$.

\item The effective polarization of neutron and proton are $P_{n} = 0.86\pm 0.02$ and $P_{p} = -0.028\pm 0.004$ taken from~\cite{Bissey:2001cw}. Similarly, the polarization of the electron beam $P_{e}$ and the polarization of the ion beam $P_{N}$ are both $70 \pm 1$\% as was stated in the EIC Yellow Report~\cite{AbdulKhalek:2021gbh}.

\item The value of $A_1^{^3\text{He}}$ can be approximated by:
\begin{equation}
    A_{1}^{^3\text{He}} =  \underbrace{{P_n\frac{F_2^n}{F_2^{^3\text{He}}}A_1^n}}_{(1)} + \underbrace{2 P_{p} \frac{F_2^p}{F_2^{^3\text{He}}} A_1^p}_{(2)},
\label{a1he}   
\end{equation}
where (1) and (2) are the neutron and the proton contributions to the asymmetry respectively. This is a simple composition model which shows how polarized $A_1^{^3\text{He}}$ can be described in terms of $A_1^n$. In reality, there are many other nuclear corrections which need to be included (ex. off-shell effects), increasing the model-dependent uncertainty.

\begin{figure*}[ht!]
    \centering
    \includegraphics[width = 0.47 \linewidth]{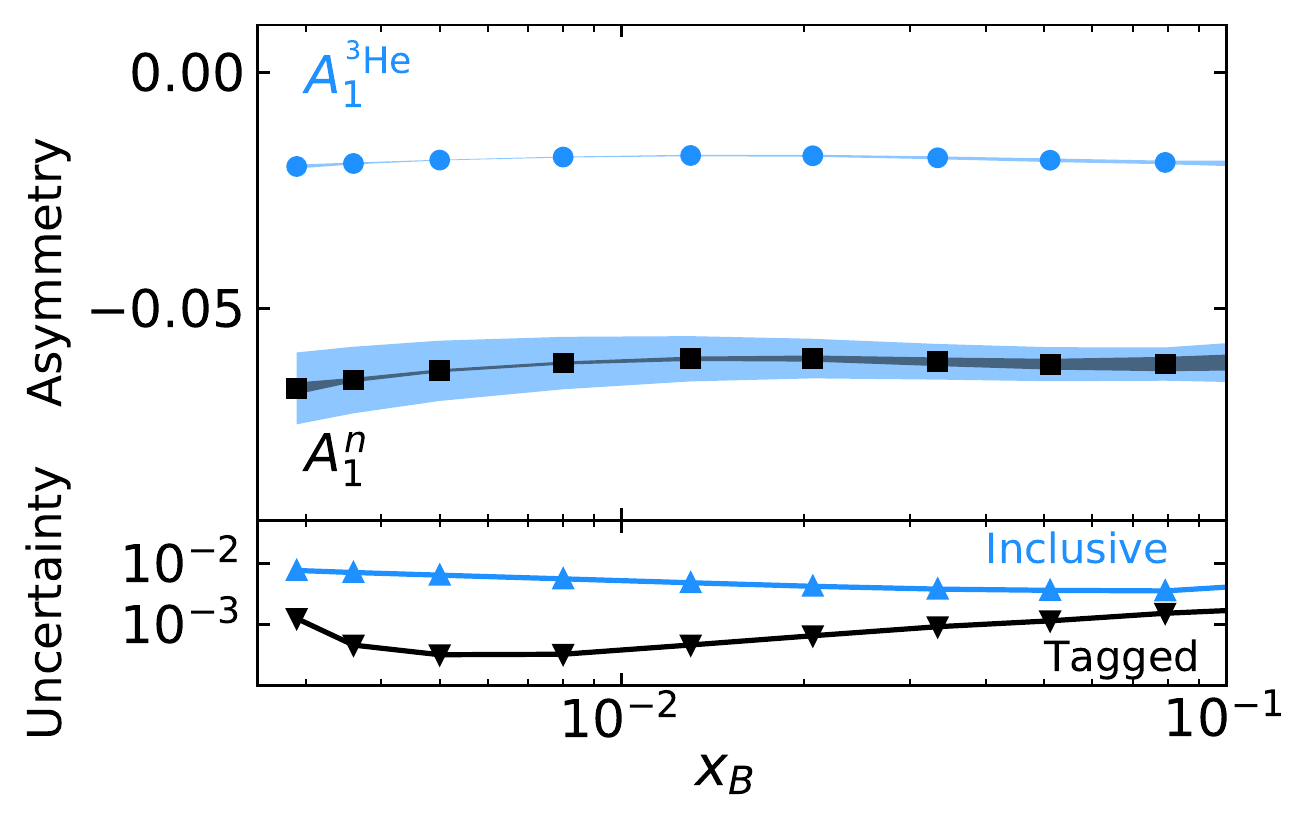}
    \includegraphics[width = 0.45 \linewidth]{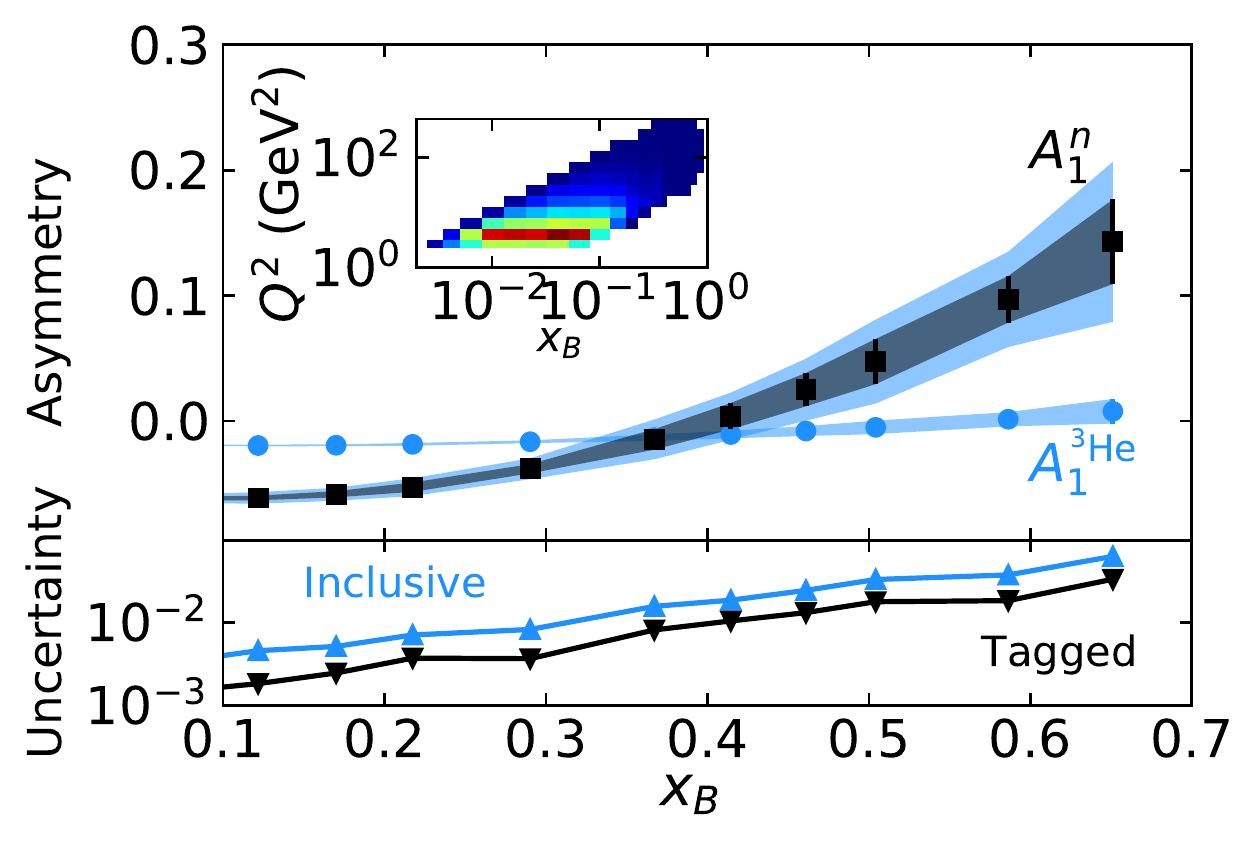}
    \caption{A direct comparison of $A^n_{1}$ extracted from inclusive measurements (blue band) and tagged measurements (black square) which are superimposed on the blue band.  The left plot is for $x \leq 0.1$ and the right plot is for $x \geq 0.1$. The blue points are the $A_1^{^3\text{He}}$ measured values from inclusive measurements from which the blue band is extracted. The uncertainties for both the techniques are compared in the bottom box where the blue (black inverted) triangles are the absolute uncertainties of inclusive (tagged) measurements. The data points were located at the average value for each $x_B$ bin. The asymmetry calculation for each data point corresponds to the average value of $Q^2$ for each $x_B$ bin.}
    \label{final}
\end{figure*}

\item In order to have good statistics in selected $x_B$ ranges, 400k of events were generated for each of three different $x_B$ ranges: $0.0026<x_B<1$, $0.2<x_B<1$ and $0.5<x_B<1$. This was repeated for unpolarized neutrons and polarized neutrons with polarizations $+1$ and $-1$. This resulted in a total of 2.4 million generated events. These events were then used to separately add SRC and 3BBU-MF regimes. As such, 4.8 million events were processed by the Eic-smear framework. Therefore, fast simulation framework Eic-smear with FF detectors was used to process these data.  

Our results are shown for integrated e-nucleon luminosity of 100 fb$^{-1}$ which corresponds to 115 days of running at 100\% efficiency with a e-nucleon luminosity of $10^{34}$ cm$^{-2}$ s$^{-1}$\cite{AbdulKhalek:2021gbh}. We assume to measure longitudinal and transverse asymmetry with the same integrated luminosity. The total events thus pass all selection cuts were binned in $x_B$. 
In this analysis, the reconstructed $x_B$ was calculated following the standard inclusive method in which only the reconstructed momentum information of the scattered electron was used\footnote{For different methods to determine $x_B$ and Q$^2$ see Ref.~\cite{Blumlein:2012bf}.}. Due to resolution effects and  a rapidly decreasing cross section, this technique is limited to  $x_B = 0.75$ for the proposed EIC detector~\cite{AbdulKhalek:2021gbh}. 
Also, for a given reconstructed $x_B$ bin, the average value of generated $x_B$ ($x$-avg) was calculated and plotted at this value~\cite{Lafferty:1994cj}. Similarly, the generated-average $Q^{2}$ ($Q^2$-avg) for a given $x_B$ was calculated.  The absolute uncertainty obtained is statistical and given by:
\begin{equation}
    \delta A_{||, \perp}^{^3\text{He}(n)} = \frac{1}{\sqrt{N}P_{e}P_{N}},
\label{abs-unc}    
\end{equation}
where $N$ is the total number of events obtained after scaling the generated events to EIC luminosity. 
The $\delta A_{1}^{^3\text{He}(n)}$ is the propagation uncertainty of $\delta A_{||, \perp}^{^3\text{He}(n)}$ through Eq.\ref{A1_formula}. 
At the high value of $Q^{2}$, both variables $\eta$ and $\xi$ are close to zero which makes the contribution of $\delta A_{\perp}^{^3\text{He}(n)}$ to $\delta A_{1}^{^3\text{He}(n)}$ small ($< 1\%)$.  Thus, for this measurement if the transverse integrated luminosity was only 10 fb$^{-1}$ it would not significantly effect the results.

\item Using the obtained value of $A_1^{^3\text{He}}$ from the previous step, $A_1^{p}$ from~\cite{E99117-PRC}, $F_2^p$ and $F_2^\text{D}$ from fit NMC E155~\cite{Arneodo:1995cq}, and $P_{p(n)}$ from~\cite{Bissey:2001cw}, we extract $A_1^{n}$ using Eq.~\ref{a1he}. The total $A_1^n$ uncertainty (shown as a blue band in Fig.~\ref{final}) is propagated from the statistical uncertainty of $A_1^{^3\text{He}}$, and systematic uncertainty from $A_1^p$, $F_2^{n}$, $F_2^\text{D}$ and $P_{p(n)}$. 

\item Double tagged events are selected using DIS cuts and requiring a condition on these two spectator protons with additional cut on $|\vec{p}_{s1} + \vec{p}_{s2}| < 0.1$ GeV/$c$ in the fixed target frame. The uncertainty of $A_1^n$ from the double tagging simulation is statistical and calculated using Eq.~\ref{abs-unc}. $F_2^n$ can also be obtained directly using the same double spectator approach but we use the value obtained from $F_2^n = F_2^\text{D} - F_2^p$ in the extraction for simplicity.

\item The $A_1^{n}$ obtained from a traditional extraction using a fixed target as well as the one obtained directly from the double spectator tagging method are shown in Fig.~\ref{final}. A comparison of the associated uncertainties using these two methods is shown in a box at the bottom of Fig.~\ref{final}.  For equal integrated luminosity, the tagged measurements provide a significant improvement on the statistical uncertainty as well as overall uncertainty from the extraction procedure, which is dominated by model dependent uncertainties from $F_{2}^{n}$, $P_{p}$ and $A^{p}_{1}$ for the fixed-target case.

\end{itemize}
The advantage of the double spectator tagging approach is that this method minimizes the uncertainty in the proton polarization, $P_{p}$ which is typically model dependent and $\sim 10\%$. While there is a good overlap ($0.4 < x_{B} < 0.65$) with the $A_{1}^{n}$ measurement conducted in Hall C at Jefferson Lab \cite{E12-06-110}, this work provides a complementary measurement at much higher $Q^2$ values thereby expanding the available world data on $A_{1}^{n}$.

\section{Radiative and Electroweak Corrections}

The effects of electromagnetic radiation on this measurement were studied using the DJANGOH event generator~\cite{Kwiatkowski:1990es,Charchula:1994kf}, which includes full radiative cross section calculations for DIS events, including the emission of radiative photons distorting the measured (leptonic) values of $x_B$ and $Q^2$ from those of the vertex (hadronic). Using the nuclear Parton Distribution Functions for $^3$He from nCTEQ15~\cite{Kovarik:2015cma}, an inclusive sample of DIS events from $^3$He was generated and the effects of experimental acceptance and resolution applied using Eic-smear. The simulated yields were compared for the case with radiative generation and the case with radiation disabled, for both the full event sample and the subset of events resulting from an initial-state neutron.

\begin{figure}[htb]
\includegraphics[width=\linewidth]{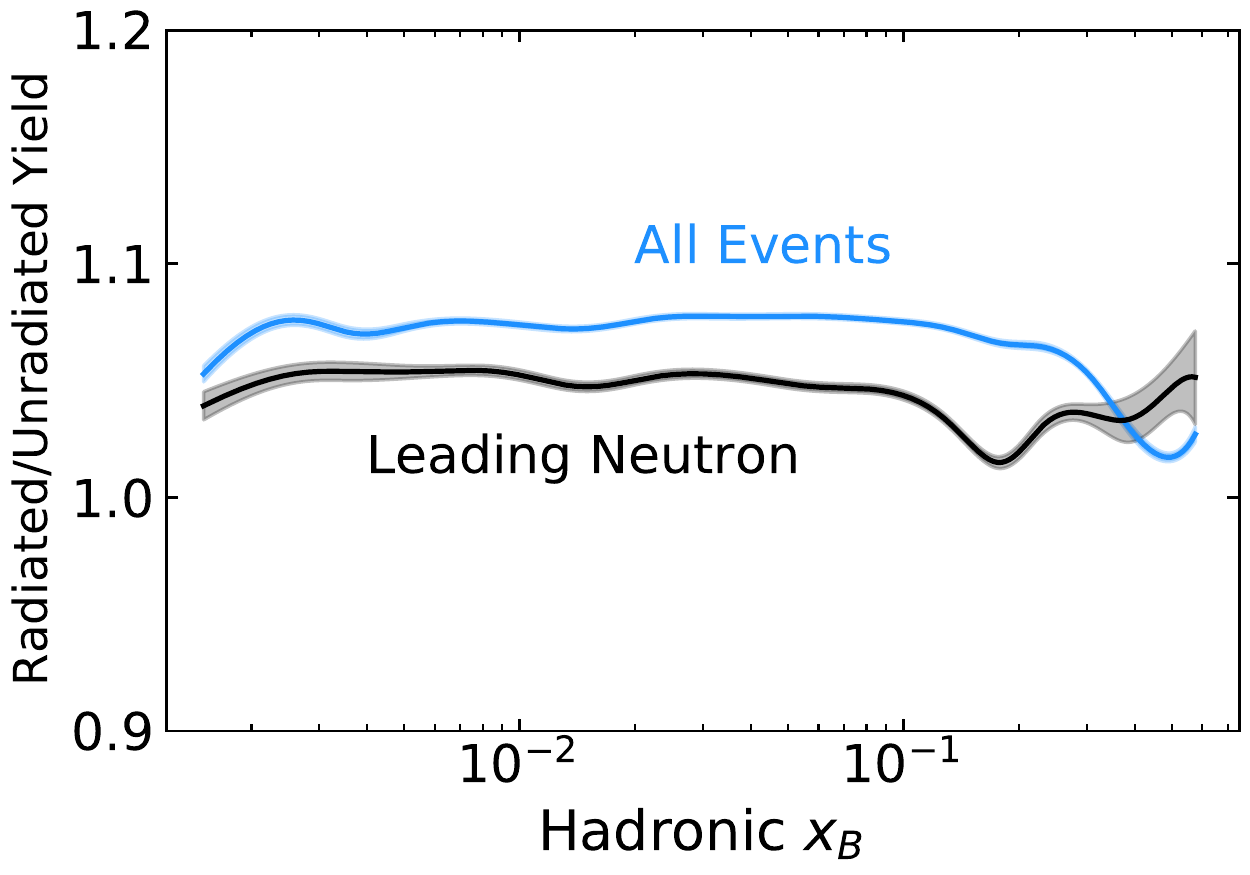}
\caption{The ratio between the simulated yield in hadronic $x_B$ for the DJANGOH samples with and without radiative effects included. The ratio is shown both for the full sample of events and for the subset of events resulting from leading neutrons, though the effects of spectator acceptance were not included in these simulations.}
\label{fig:radiation}
\end{figure}

The primary effect of single-photon emission was found to be to lower the leptonic value of $x_B$ from that of the interaction, as well as to smear the leptonic value of $Q^2$. This has an apparently large effect on the measured distributions, particularly inflating the yield in the lowest-$x_B$ kinematic bins. However, the primary concern for measurements of $A_1^n$ is the coverage in the hadronic kinematic variables. The effect of radiation on the coverage of hadronic $x_B$, shown in Fig.~\ref{fig:radiation}, was found to be significantly smaller, comprising a $<10$\% impact on yield over the measured bins and therefore a $<5$\% impact on statistical precision. This was found to be the case both for the full set of simulated DIS events and for those events which resulted from a leading neutron, indicating a similarly small effect for both the inclusive and tagged measurements.

In an analysis of this data, which would include such radiative effects, it would be necessary to compare with physics models such as DJANGOH which fully calculate the impact of radiation on the interaction, with the addition of the spectator system for fully tagged events. By including possible models for $A_1$ in the event generation and using realistic detector models, all physics and experimental effects can be convoluted into pseudodata for comparison with measured data, both for the inclusive and tagged case. This method avoids possible model-dependencies in the $A_1^n$ extraction which would result from attempting to correct data for radiative effects and would allow inference on fit parameters for the model of $A_1^n$.

Finally, while the simulations in this work use pure  electromagnetic structure functions, real data in these EIC kinematics will have electroweak contributions to the asymmetry which must be considered.  For elastic scattering, the electroweak (\textit{i.e.} parity-violating) contribution to the asymmetry is as large as $10^{-4}$~\cite{Sofiatti:2011yi}, much smaller than the $1\%$ asymmetry found herein for $\rm ^3{He}$.  Elastic scattering decreases with $Q^2$ much faster than DIS, which to first order is constant in $Q^2$; thus, the parity-violating contribution of electromagnetic DIS/SIDIS due to weak interactions is expected to be a significant correction for precise measurements only in certain kinematics, \textit{e.g.} for large $Q^2$.  Consideration of parity-violating asymmetries will require their calculation bin-by-bin in the kinematics of the experiment.  

%Additionally, by averaging over the polarization of the ion the asymmetry of the electroweak to electromagnetic ratio is reduced.  Therefore, for an experiment on polarized $^3$He at the EIC, it is desirable to have a complete set of measurements of both transverse and longitudinal polarization in order to average independently over electron and ion spins as well as transverse plus longitudinal. This will experimentally allow scientists to isolate the electroweak and electromagnetic contributions and $A_1$ and $A_2$. The Q$^2$ dependence can be used as a further cross-check that these extractions have been done correctly.

\section{Conclusion}
We have shown that with the advent of the EIC, neutron structure can be investigated through the unique far-forward detection capabilities to tag double spectator protons -- remnants of electron scattering events from the neutron in polarized $^3$He ions. 
This double spectator tagging technique greatly reduces the systematic uncertainties thereby allowing an extraction of $A^n_{1}$ in a precise and model independent manner while simultaneously determining the validity of multiple theoretical models used in $A^n_{1}$ extraction for fixed target scenarios. 
Improvement in precision in $A^n_{1}$ as compared to inclusive extractions is found to be a factor of $3$ over all kinematics and greater than an order-of-magnitude in the low-$x_B$ region.
This novel approach provides crucial new information to our understanding of the role of valence quarks in neutron spin structure with far less model dependence than other techniques, and will help to reveal possible quark orbital angular momentum contributions to the spin structure of the neutron.

\section*{Acknowledgements}

This  work  was  funded  in  part  by  the US Department of Energy contracts DE-SC0020240, DE-SC0012704,
and also DE-AC05-06OR23177, under which Jefferson Science Associates, LLC operates the Thomas Jefferson National Accelerator Facility. We also acknowledge the support of the Jefferson Lab EIC Center. The work of A. Jentsch was further supported by the Program Development program at Brookhaven National Laboratory. The work of Z.Tu is supported by LDRD-039 and the Goldhaber Distinguished Fellowship at Brookhaven National Laboratory. J.R.W's work is partially supported by the LDRD program of Lawrence Berkeley National Laboratory, the U.S. Department of Energy, Office of Science, Office of Nuclear Physics, under contract number DE-AC02-05CH11231, and within the framework of the TMD Topical Collaboration.
The work of I.~Fri\v{s}\v{c}i\'{c}, J.R.~Pybus, and J.R.W was supported by EIC Center fellowships at Jefferson Lab.

\bibliography{references}

\begin{thebibliography}{10}
\expandafter\ifx\csname url\endcsname\relax
  \def\url#1{\texttt{#1}}\fi
\expandafter\ifx\csname urlprefix\endcsname\relax\def\urlprefix{URL }\fi
\expandafter\ifx\csname href\endcsname\relax
  \def\href#1#2{#2} \def\path#1{#1}\fi

\bibitem{Kuhn:2008sy}
S.~E. Kuhn, J.~P. Chen, E.~Leader, {Spin Structure of the Nucleon - Status and
  Recent Results}, Prog. Part. Nucl. Phys. 63 (2009) 1--50.
\newblock \href {http://arxiv.org/abs/0812.3535} {\path{arXiv:0812.3535}},
  \href {https://doi.org/10.1016/j.ppnp.2009.02.001}
  {\path{doi:10.1016/j.ppnp.2009.02.001}}.

\bibitem{Leader:2013jra}
E.~Leader, C.~Lorc\'e, {The angular momentum controversy:
  What\textquoteright{}s it all about and does it matter?}, Phys. Rept. 541~(3)
  (2014) 163--248.
\newblock \href {http://arxiv.org/abs/1309.4235} {\path{arXiv:1309.4235}},
  \href {https://doi.org/10.1016/j.physrep.2014.02.010}
  {\path{doi:10.1016/j.physrep.2014.02.010}}.

\bibitem{1988PhLB..206..364A}
J.~{Ashman}, {European Muon Collaboration}, {A measurement of the spin
  asymmetry and determination of the structure function g$_{1}$ in deep
  inelastic muon-proton scattering}, Physics Letters B 206~(2) (1988) 364--370.
\newblock \href {https://doi.org/10.1016/0370-2693(88)91523-7}
  {\path{doi:10.1016/0370-2693(88)91523-7}}.

\bibitem{RevModPhys.85.655}
C.~A. Aidala, S.~D. Bass, D.~Hasch, G.~K. Mallot,
  \href{https://link.aps.org/doi/10.1103/RevModPhys.85.655}{The spin structure
  of the nucleon}, Rev. Mod. Phys. 85 (2013) 655--691.
\newblock \href {https://doi.org/10.1103/RevModPhys.85.655}
  {\path{doi:10.1103/RevModPhys.85.655}}.
\newline\urlprefix\url{https://link.aps.org/doi/10.1103/RevModPhys.85.655}

\bibitem{Ji:2020ena}
X.~Ji, F.~Yuan, Y.~Zhao, {What we know and what we don\textquoteright{}t know
  about the proton spin after 30 years}, Nature Rev. Phys. 3~(1) (2021) 27--38.
\newblock \href {http://arxiv.org/abs/2009.01291} {\path{arXiv:2009.01291}},
  \href {https://doi.org/10.1038/s42254-020-00248-4}
  {\path{doi:10.1038/s42254-020-00248-4}}.

\bibitem{E99117-PRL}
X.~Zheng, et~al., {Precision measurement of the neutron spin asymmetry A1**N
  and spin flavor decomposition in the valence quark region}, Phys. Rev. Lett.
  92 (2004) 012004.
\newblock \href {http://arxiv.org/abs/nucl-ex/0308011}
  {\path{arXiv:nucl-ex/0308011}}, \href
  {https://doi.org/10.1103/PhysRevLett.92.012004}
  {\path{doi:10.1103/PhysRevLett.92.012004}}.

\bibitem{Freese:2020mcx}
A.~Freese, I.~C. Clo\"et, {Quark spin and orbital angular momentum from proton
  generalized parton distributions}, Phys. Rev. C 103~(4) (2021) 045204.
\newblock \href {http://arxiv.org/abs/2005.10286} {\path{arXiv:2005.10286}},
  \href {https://doi.org/10.1103/PhysRevC.103.045204}
  {\path{doi:10.1103/PhysRevC.103.045204}}.

\bibitem{Deur:2018roz}
A.~Deur, S.~J. Brodsky, G.~F. De~T\'eramond, {The Spin Structure of the
  Nucleon}, Rept.Prog.Phys. 82 (2019) 076201.
\newblock \href {http://arxiv.org/abs/1807.05250} {\path{arXiv:1807.05250}},
  \href {https://doi.org/10.1088/1361-6633/ab0b8f}
  {\path{doi:10.1088/1361-6633/ab0b8f}}.

\bibitem{COHERENT:2015mry}
D.~Akimov, et~al., {The COHERENT Experiment at the Spallation Neutron Source}
  (2015).
\newblock \href {http://arxiv.org/abs/1509.08702} {\path{arXiv:1509.08702}}.

\bibitem{Bodek:1979rx}
A.~Bodek, et~al., {Experimental Studies of the Neutron and Proton
  Electromagnetic Structure Functions}, Phys. Rev. D 20 (1979) 1471--1552.
\newblock \href {https://doi.org/10.1103/PhysRevD.20.1471}
  {\path{doi:10.1103/PhysRevD.20.1471}}.

\bibitem{Arrington:2008zh}
J.~Arrington, F.~Coester, R.~J. Holt, T.~S.~H. Lee, {Neutron Structure
  Functions}, J. Phys. G 36 (2009) 025005.
\newblock \href {http://arxiv.org/abs/0805.3116} {\path{arXiv:0805.3116}},
  \href {https://doi.org/10.1088/0954-3899/36/2/025005}
  {\path{doi:10.1088/0954-3899/36/2/025005}}.

\bibitem{Anthony:1993uf}
P.~L. Anthony, et~al., {Determination of the neutron spin structure function.},
  Phys. Rev. Lett. 71 (1993) 959--962.
\newblock \href {https://doi.org/10.1103/PhysRevLett.71.959}
  {\path{doi:10.1103/PhysRevLett.71.959}}.

\bibitem{Sulkosky:2019zmn}
V.~Sulkosky, et~al., {Measurement of the $^3He$ spin-structure functions and of
  neutron ($^3$He) spin-dependent sum rules at $0.035\le Q^2 \le 0.24$
  GeV$^2$}, Phys. Lett. B 805 (2020) 135428.
\newblock \href {http://arxiv.org/abs/1908.05709} {\path{arXiv:1908.05709}},
  \href {https://doi.org/10.1016/j.physletb.2020.135428}
  {\path{doi:10.1016/j.physletb.2020.135428}}.

\bibitem{E99117-PRC}
X.~Zheng, et~al., {Precision measurement of the neutron spin asymmetries and
  spin-dependent structure functions in the valence quark region}, Phys. Rev. C
  70 (2004) 065207.
\newblock \href {http://arxiv.org/abs/nucl-ex/0405006}
  {\path{arXiv:nucl-ex/0405006}}, \href
  {https://doi.org/10.1103/PhysRevC.70.065207}
  {\path{doi:10.1103/PhysRevC.70.065207}}.

\bibitem{EICCDR}
J.~A. et~al., \href{https://www.bnl.gov/ec/files/EIC_CDR_Final.pdf}{Electron
  ion collider conceptual design report} (2021).
\newline\urlprefix\url{https://www.bnl.gov/ec/files/EIC_CDR_Final.pdf}

\bibitem{Maxwell:2016knr}
J.~Maxwell, C.~Epstein, R.~Milner, J.~Alessi, E.~Beebe, A.~Pikin, J.~Ritter,
  A.~Zelenski, {Development of a Polarized Helium-3 Source for RHIC and eRHIC},
  Int. J. Mod. Phys. Conf. Ser. 40~(01) (2016) 1660102.
\newblock \href {https://doi.org/10.1142/S2010194516601022}
  {\path{doi:10.1142/S2010194516601022}}.

\bibitem{Baillie:2011za}
N.~Baillie, et~al., {Measurement of the neutron F2 structure function via
  spectator tagging with CLAS}, Phys. Rev. Lett. 108 (2012) 142001, [Erratum:
  Phys.Rev.Lett. 108, 199902 (2012)].
\newblock \href {http://arxiv.org/abs/1110.2770} {\path{arXiv:1110.2770}},
  \href {https://doi.org/10.1103/PhysRevLett.108.142001}
  {\path{doi:10.1103/PhysRevLett.108.142001}}.

\bibitem{Tkachenko:2014byy}
S.~Tkachenko, et~al., {Measurement of the structure function of the nearly free
  neutron using spectator tagging in inelastic $^2$H(e, e'p)X scattering with
  CLAS}, Phys. Rev. C 89 (2014) 045206, [Addendum: Phys.Rev.C 90, 059901
  (2014)].
\newblock \href {http://arxiv.org/abs/1402.2477} {\path{arXiv:1402.2477}},
  \href {https://doi.org/10.1103/PhysRevC.89.045206}
  {\path{doi:10.1103/PhysRevC.89.045206}}.

\bibitem{Cosyn:2014zfa}
W.~Cosyn, V.~Guzey, D.~W. Higinbotham, C.~Hyde, S.~Kuhn, P.~Nadel-Turonski,
  K.~Park, M.~Sargsian, M.~Strikman, C.~Weiss, {Neutron spin structure with
  polarized deuterons and spectator proton tagging at EIC}, J. Phys. Conf. Ser.
  543 (2014) 012007.
\newblock \href {http://arxiv.org/abs/1409.5768} {\path{arXiv:1409.5768}},
  \href {https://doi.org/10.1088/1742-6596/543/1/012007}
  {\path{doi:10.1088/1742-6596/543/1/012007}}.

\bibitem{Cosyn:2019hem}
W.~Cosyn, C.~Weiss, {Neutron spin structure from polarized deuteron DIS with
  proton tagging}, Phys. Lett. B 799 (2019) 135035.
\newblock \href {http://arxiv.org/abs/1906.11119} {\path{arXiv:1906.11119}},
  \href {https://doi.org/10.1016/j.physletb.2019.135035}
  {\path{doi:10.1016/j.physletb.2019.135035}}.

\bibitem{Cosyn:2020kwu}
W.~Cosyn, C.~Weiss, {Polarized electron-deuteron deep-inelastic scattering with
  spectator nucleon tagging}, Phys. Rev. C 102 (2020) 065204.
\newblock \href {http://arxiv.org/abs/2006.03033} {\path{arXiv:2006.03033}},
  \href {https://doi.org/10.1103/PhysRevC.102.065204}
  {\path{doi:10.1103/PhysRevC.102.065204}}.

\bibitem{Flay:2016wie}
D.~Flay, et~al., {Measurements of $d_{2}^{n}$ and $A_{1}^{n}$: Probing the
  neutron spin structure}, Phys. Rev. D 94~(5) (2016) 052003.
\newblock \href {http://arxiv.org/abs/1603.03612} {\path{arXiv:1603.03612}},
  \href {https://doi.org/10.1103/PhysRevD.94.052003}
  {\path{doi:10.1103/PhysRevD.94.052003}}.

\bibitem{Mihovilovic:2018fux}
M.~Mihovilovi\v{c}, et~al., {Measurement of double-polarization asymmetries in
  the quasi-elastic
  $^3\vec{\mathrm{He}}(\vec{\mathrm{e}},\mathrm{e}'\mathrm{p})$ process}, Phys.
  Lett. B 788 (2019) 117--121.
\newblock \href {http://arxiv.org/abs/1804.06043} {\path{arXiv:1804.06043}},
  \href {https://doi.org/10.1016/j.physletb.2018.10.063}
  {\path{doi:10.1016/j.physletb.2018.10.063}}.

\bibitem{Mihovilovic:2014gdi}
M.~Mihovilovic, et~al., {Measurement of double-polarization asymmetries in the
  quasielastic $^3\vec{\mathrm{He}}(\vec{\mathrm{e}},\mathrm{e}'\mathrm{d})$
  process}, Phys. Rev. Lett. 113~(23) (2014) 232505.
\newblock \href {http://arxiv.org/abs/1409.2253} {\path{arXiv:1409.2253}},
  \href {https://doi.org/10.1103/PhysRevLett.113.232505}
  {\path{doi:10.1103/PhysRevLett.113.232505}}.

\bibitem{Long:2019iig}
E.~Long, et~al., {Measurement of the single-spin asymmetry $A_y^0$ in
  quasi-elastic $^3$He$^\uparrow$($e,e'n$) scattering at $0.4 < Q^2 < 1.0$
  GeV$/c^2$}, Phys. Lett. B 797 (2019) 134875.
\newblock \href {http://arxiv.org/abs/1906.04075} {\path{arXiv:1906.04075}},
  \href {https://doi.org/10.1016/j.physletb.2019.134875}
  {\path{doi:10.1016/j.physletb.2019.134875}}.

\bibitem{Cruz-Torres:2020uke}
R.~Cruz-Torres, et~al., {Probing Few-Body Nuclear Dynamics via $^3$H and $^3$He
  ($e,e'p$)pn Cross-Section Measurements}, Phys. Rev. Lett. 124~(21) (2020)
  212501.
\newblock \href {http://arxiv.org/abs/2001.07230} {\path{arXiv:2001.07230}},
  \href {https://doi.org/10.1103/PhysRevLett.124.212501}
  {\path{doi:10.1103/PhysRevLett.124.212501}}.

\bibitem{Anthony:2000fn}
P.~L. Anthony, et~al., {Measurements of the Q**2 dependence of the proton and
  neutron spin structure functions g(1)**p and g(1)**n}, Phys. Lett. B 493
  (2000) 19--28.
\newblock \href {http://arxiv.org/abs/hep-ph/0007248}
  {\path{arXiv:hep-ph/0007248}}, \href
  {https://doi.org/10.1016/S0370-2693(00)01014-5}
  {\path{doi:10.1016/S0370-2693(00)01014-5}}.

\bibitem{Friedman:1972sy}
J.~I. Friedman, H.~W. Kendall, {Deep inelastic electron scattering}, Ann. Rev.
  Nucl. Part. Sci. 22 (1972) 203--254.
\newblock \href {https://doi.org/10.1146/annurev.ns.22.120172.001223}
  {\path{doi:10.1146/annurev.ns.22.120172.001223}}.

\bibitem{Mankiewicz:1991dp}
L.~Mankiewicz, A.~Schafer, M.~Veltri, {PEPSI: A Monte Carlo generator for
  polarized leptoproduction}, Comput. Phys. Commun. 71 (1992) 305--318.
\newblock \href {https://doi.org/10.1016/0010-4655(92)90016-R}
  {\path{doi:10.1016/0010-4655(92)90016-R}}.

\bibitem{CiofidegliAtti:2005qt}
C.~Ciofi~degli Atti, L.~Kaptari, {On the interpretation of the processes
  He-3(e,e-prime p) H-2 and He-3(e,e-prime p)(pn) at high missing momenta},
  Phys. Rev. Lett. 95 (2005) 052502.
\newblock \href {http://arxiv.org/abs/nucl-th/0502045}
  {\path{arXiv:nucl-th/0502045}}, \href
  {https://doi.org/10.1103/PhysRevLett.95.052502}
  {\path{doi:10.1103/PhysRevLett.95.052502}}.

\bibitem{Pybus:2020itv}
J.~Pybus, I.~Korover, R.~Weiss, A.~Schmidt, N.~Barnea, D.~Higinbotham,
  E.~Piasetzky, M.~Strikman, L.~Weinstein, O.~Hen, {Generalized contact
  formalism analysis of the $^4$He$(e,e'pN)$ reaction}, Phys. Lett. B 805
  (2020) 135429.
\newblock \href {http://arxiv.org/abs/2003.02318} {\path{arXiv:2003.02318}},
  \href {https://doi.org/10.1016/j.physletb.2020.135429}
  {\path{doi:10.1016/j.physletb.2020.135429}}.

\bibitem{Cruz-Torres:2019fum}
R.~Cruz-Torres, D.~Lonardoni, R.~Weiss, N.~Barnea, D.~Higinbotham,
  E.~Piasetzky, A.~Schmidt, L.~Weinstein, R.~Wiringa, O.~Hen, {Scale and Scheme
  Independence and Position-Momentum Equivalence of Nuclear Short-Range
  Correlations}, Nature Physics (2020).
\newblock \href {https://doi.org/10.1038/s41567-020-01053-7}
  {\path{doi:10.1038/s41567-020-01053-7}}.

\bibitem{Frankfurt:1981mk}
L.~Frankfurt, M.~Strikman, {High-Energy Phenomena, Short Range Nuclear
  Structure and QCD}, Phys. Rept. 76 (1981) 215--347.
\newblock \href {https://doi.org/10.1016/0370-1573(81)90129-0}
  {\path{doi:10.1016/0370-1573(81)90129-0}}.

\bibitem{GEANT4}
S.~Agostinelli, et~al., {GEANT4}---a simulation toolkit, Nucl. Instrum. Meth. A
  506 (2003) 250.
\newblock \href {https://doi.org/10.1016/S0168-9002(03)01368-8}
  {\path{doi:10.1016/S0168-9002(03)01368-8}}.

\bibitem{AbdulKhalek:2021gbh}
R.~Abdul~Khalek, et~al., {Science Requirements and Detector Concepts for the
  Electron-Ion Collider: EIC Yellow Report} (3 2021).
\newblock \href {http://arxiv.org/abs/2103.05419} {\path{arXiv:2103.05419}}.

\bibitem{Tu:2020BeAGLE}
Z.~Tu, A.~Jentsch, M.~Baker, L.~Zheng, J.~H. Lee, R.~Venugopalan, O.~Hen,
  D.~Higinbotham, E.~C. Aschenauer, T.~Ullrich, {Probing Short-Range
  Correlations in the Deuteron via Incoherent Diffractive J/Psi Production with
  Spectator Tagging at the EIC}, Phys. Lett. B 811 (2020) 135877.
\newblock \href {https://doi.org/10.1016/j.physletb.2020.135877}
  {\path{doi:10.1016/j.physletb.2020.135877}}.

\bibitem{matrix02}
\url{https://eic.github.io/software/eicsmear.html#matrix-02} (2021).

\bibitem{Sato:2016tuz}
N.~Sato, W.~Melnitchouk, S.~E. Kuhn, J.~J. Ethier, A.~Accardi, {Iterative Monte
  Carlo analysis of spin-dependent parton distributions}, Phys. Rev. D 93~(7)
  (2016) 074005.
\newblock \href {http://arxiv.org/abs/1601.07782} {\path{arXiv:1601.07782}},
  \href {https://doi.org/10.1103/PhysRevD.93.074005}
  {\path{doi:10.1103/PhysRevD.93.074005}}.

\bibitem{Feynman:1973xc}
R.~P. Feynman, {Photon-hadron interactions} (1973).

\bibitem{E143:1998nvx}
K.~Abe, et~al., {Measurements of R = $\sigma$(L) / $\sigma$(T) for 0.03
  \ensuremath{<} x \ensuremath{<} 0.1 and fit to world data}, Phys. Lett. B 452
  (1999) 194--200.
\newblock \href {http://arxiv.org/abs/hep-ex/9808028}
  {\path{arXiv:hep-ex/9808028}}, \href
  {https://doi.org/10.1016/S0370-2693(99)00244-0}
  {\path{doi:10.1016/S0370-2693(99)00244-0}}.

\bibitem{Arneodo:1995cq}
M.~Arneodo, et~al., {Measurement of the proton and the deuteron structure
  functions, F2(p) and F2(d)}, Phys. Lett. B 364 (1995) 107--115.
\newblock \href {http://arxiv.org/abs/hep-ph/9509406}
  {\path{arXiv:hep-ph/9509406}}, \href
  {https://doi.org/10.1016/0370-2693(95)01318-9}
  {\path{doi:10.1016/0370-2693(95)01318-9}}.

\bibitem{Bissey:2001cw}
F.~R.~P. Bissey, V.~A. Guzey, M.~Strikman, A.~W. Thomas, {Complete analysis of
  spin structure function g(1) of He-3}, Phys. Rev. C 65 (2002) 064317.
\newblock \href {http://arxiv.org/abs/hep-ph/0109069}
  {\path{arXiv:hep-ph/0109069}}, \href
  {https://doi.org/10.1103/PhysRevC.65.064317}
  {\path{doi:10.1103/PhysRevC.65.064317}}.

\bibitem{Blumlein:2012bf}
J.~Blumlein, {The Theory of Deeply Inelastic Scattering}, Prog. Part. Nucl.
  Phys. 69 (2013) 28--84.
\newblock \href {http://arxiv.org/abs/1208.6087} {\path{arXiv:1208.6087}},
  \href {https://doi.org/10.1016/j.ppnp.2012.09.006}
  {\path{doi:10.1016/j.ppnp.2012.09.006}}.

\bibitem{Lafferty:1994cj}
G.~D. Lafferty, T.~R. Wyatt, {Where to stick your data points: The treatment of
  measurements within wide bins}, Nucl. Instrum. Meth. A 355 (1995) 541--547.
\newblock \href {https://doi.org/10.1016/0168-9002(94)01112-5}
  {\path{doi:10.1016/0168-9002(94)01112-5}}.

\bibitem{E12-06-110}
X.~Zheng, G.~Cates, J.~Chen, Z.~Meziani, {Measurement of neutron spin asymmetry
  $A^{n}_{1}$ in the valence quark region using an 11 GeV beam and a Polarized
  $^3$He target in Hall C}, proposal E12-06-110 (2006).

\bibitem{Kwiatkowski:1990es}
A.~Kwiatkowski, H.~Spiesberger, H.~J. Mohring, {Heracles: An Event Generator
  for $e p$ Interactions at {HERA} Energies Including Radiative Processes:
  Version 1.0}, Comput. Phys. Commun. 69 (1992) 155--172.
\newblock \href {https://doi.org/10.1016/0010-4655(92)90136-M}
  {\path{doi:10.1016/0010-4655(92)90136-M}}.

\bibitem{Charchula:1994kf}
K.~Charchula, G.~A. Schuler, H.~Spiesberger, {Combined QED and QCD radiative
  effects in deep inelastic lepton - proton scattering: The Monte Carlo
  generator DJANGO6}, Comput. Phys. Commun. 81 (1994) 381--402.
\newblock \href {https://doi.org/10.1016/0010-4655(94)90086-8}
  {\path{doi:10.1016/0010-4655(94)90086-8}}.

\bibitem{Kovarik:2015cma}
K.~Kovarik, et~al., {nCTEQ15 - Global analysis of nuclear parton distributions
  with uncertainties in the CTEQ framework}, Phys. Rev. D 93~(8) (2016) 085037.
\newblock \href {http://arxiv.org/abs/1509.00792} {\path{arXiv:1509.00792}},
  \href {https://doi.org/10.1103/PhysRevD.93.085037}
  {\path{doi:10.1103/PhysRevD.93.085037}}.

\bibitem{Sofiatti:2011yi}
C.~Sofiatti, T.~W. Donnelly, {Polarized e-p Elastic Scattering in the Collider
  Frame}, Phys. Rev. C 84 (2011) 014606.
\newblock \href {http://arxiv.org/abs/1104.2149} {\path{arXiv:1104.2149}},
  \href {https://doi.org/10.1103/PhysRevC.84.014606}
  {\path{doi:10.1103/PhysRevC.84.014606}}.

\end{thebibliography}

\end{document}

\title{Supplementary Materials: }
\pacs{}
\maketitle

\section{Fermi-correction} %Jackson

The process for generating tagged $^3\text{He}(e,e'pp)X$ events is described here. 
The CLASDIS event generator was used to produce a sample of $n(e,e')X$ DIS events from a stationary neutron target, including the scattered electron momentum and the hadronic remnant for each event. 
Each event was assigned an initial nuclear state; the nuclear state was sampled from the light-front spectral function of $^3\text{He}$, the model for which is described in the main text. 
For each nuclear state, the system is described by:
\begin{itemize}
    \item The momentum perpendicular to the momentum transfer $\bf p^\perp_i$
    \item The light-front momentum fraction $\alpha_i=\frac{A}{m_A}p_i^-=\frac{A}{m_A}(E_i-p_i^z)$ where $A$ is the atomic mass number and $m_A$ is the target mass and the $z$-component is defined parallel to the momentum transfer.
\end{itemize}
Here $i=1,s1,s2$ denotes the struck neutron and the spectator protons, respectively.

The DIS event on a nucleon at rest is defined by the initial four-momentum of the electron $k$, the final four-momentum of the electron $k'$ and the four-momentum transfer $q=k-k'$.
We define the following scalar quantities for the event in the target rest frame:
\begin{itemize}
  \item The Bjorken variable $x_B=\frac{Q^2}{2m_N\omega}$, where $Q^2=-q^2$ is the squared four-momentum transfer, $m_{N}$ is the mass of the nucleon, and $\omega=E_{e}-E'_{e}$ is the energy transfer in the target rest frame
  \item The fractional energy transfer $y=\frac{\omega}{E_{e}}$
  \item The final hadronic system squared invariant mass $W^2=m_N^2 + 2m_N\omega - Q^2$, where the target is assumed to be a nucleon at rest
  \item The energy of the beam electron $E_{e}$
  \item The energy $E_e'$, opening angle $\theta_e'$, and azimuthal angle $\phi_e'$ of the scattered electron
\end{itemize}

The variables $x_B$, $\omega$, and $\phi_e'$ are calculated for a given event, and fully constrain the DIS event from a nucleon at rest. The nuclear initial state is sampled for the event, and the variables $\bf p^\perp_i$ and $\alpha_i$ are determined for each nucleon $i$. For the spectator nucleons $i=s1,s2$, these variables are used to calculate the four-momentum of the spectator nucleons in the final state. For the leading nucleon $i=1$, these variables are used to alter the electron kinematics to account for the nuclear motion. 

To perform this Fermi correction, the kinematic variable $x_B$ is multiplied by a factor of the leading nucleon lightcone fraction:
\begin{equation}
    x_{B,new}=\alpha x_B
\end{equation}
The variables $\omega$ and $\phi_e'$ remain unchanged: 
\begin{gather}
    \omega_{new}=\omega \\
    \phi_{e,new}'=\phi_e'
\end{gather}
This provides a new set of scattering variables to fully constrain the electron 4-momentum. These variables are used to recalculate the electron momentum:
\begin{gather}
    E_{e,new}' = E_e - \omega_{new} \\
    Q^2_{new}=2m_N\omega_{new} x_{B,new} \\
    \cos \theta_{e,new}'=1\frac{Q^2_{new}}{2E_eE_{e,new}'}
\end{gather}
This provides the full information of the final electron four-momentum $k'_{new}$. From this, the Fermi-corrected four-momentum transfer $q_{new}$ is calculated. Finally, the momentum of the spectator nucleons, which has to this point been defined only relative to the $\bf q$ vector, is rotated to the appropriate orientation relative to the momentum transfer. This provides the four-momentum of the final state electron and spectator nucleons event-by-event for use in simulation.

\section{kinematic plots} %Dien

\begin{figure*}[htb]
    \centering
    \includegraphics[width=0.48\linewidth]{Figures/EIC_root_plots/pmiss_5x41.pdf}
    \includegraphics[width=0.48\linewidth]{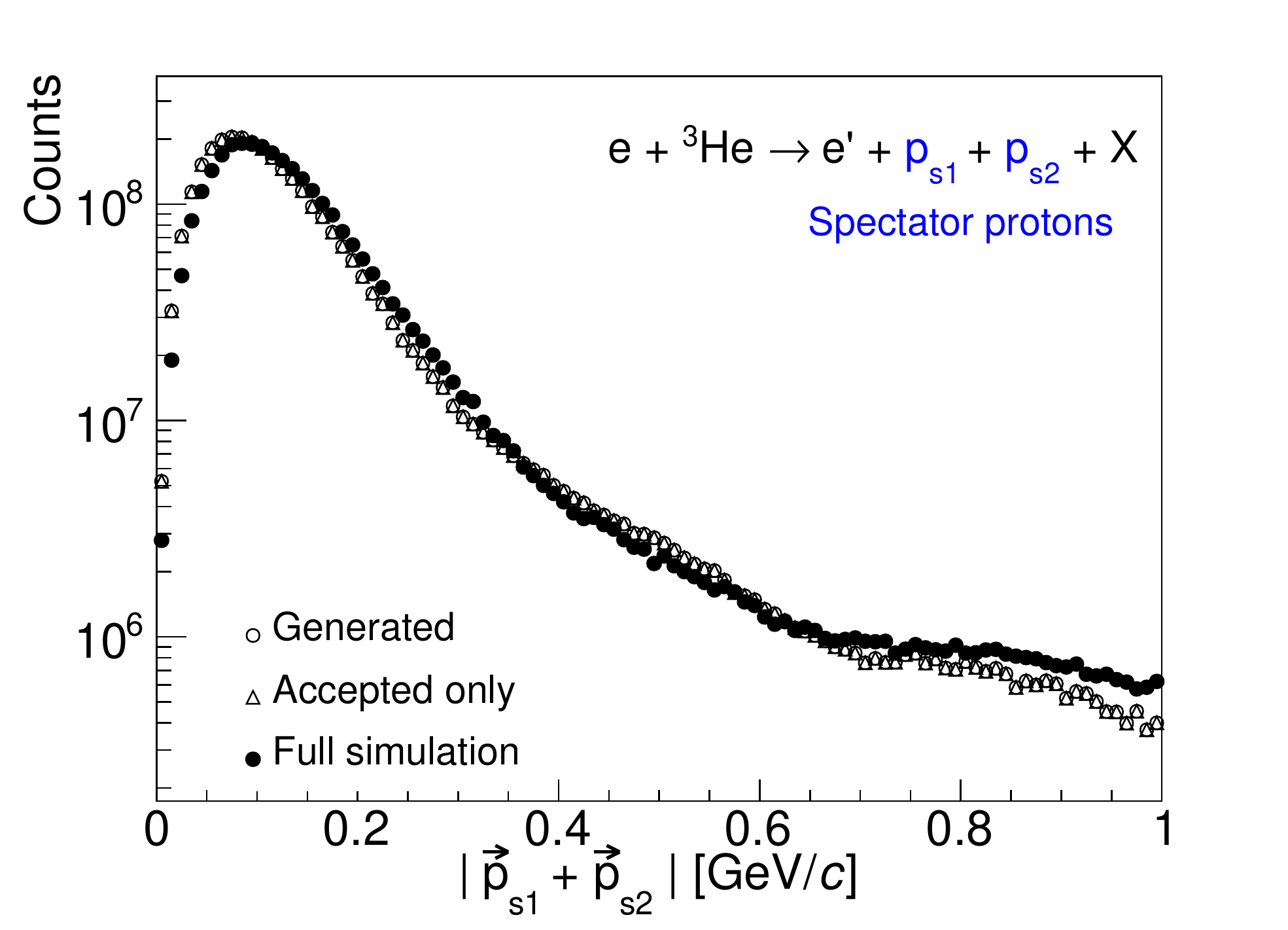}
    \caption{Distribution of the sum of two spectator protons, $p_{s1}$ and $p_{s2}$ in Fig.~\ref{fig:diagram}, in the Ion Rest Frame (IRF) for the kinematic setting 5x41 GeV on the left and 18x110 GeV on the right. The simulation at the generated level, with acceptance correction and for the full simulation was shown in open circles, open triangles and the solid points, respectively.}
    \label{He3-41x5}
\end{figure*}

\begin{figure}[ht]
    \centering
    \includegraphics[width=0\linewidth]{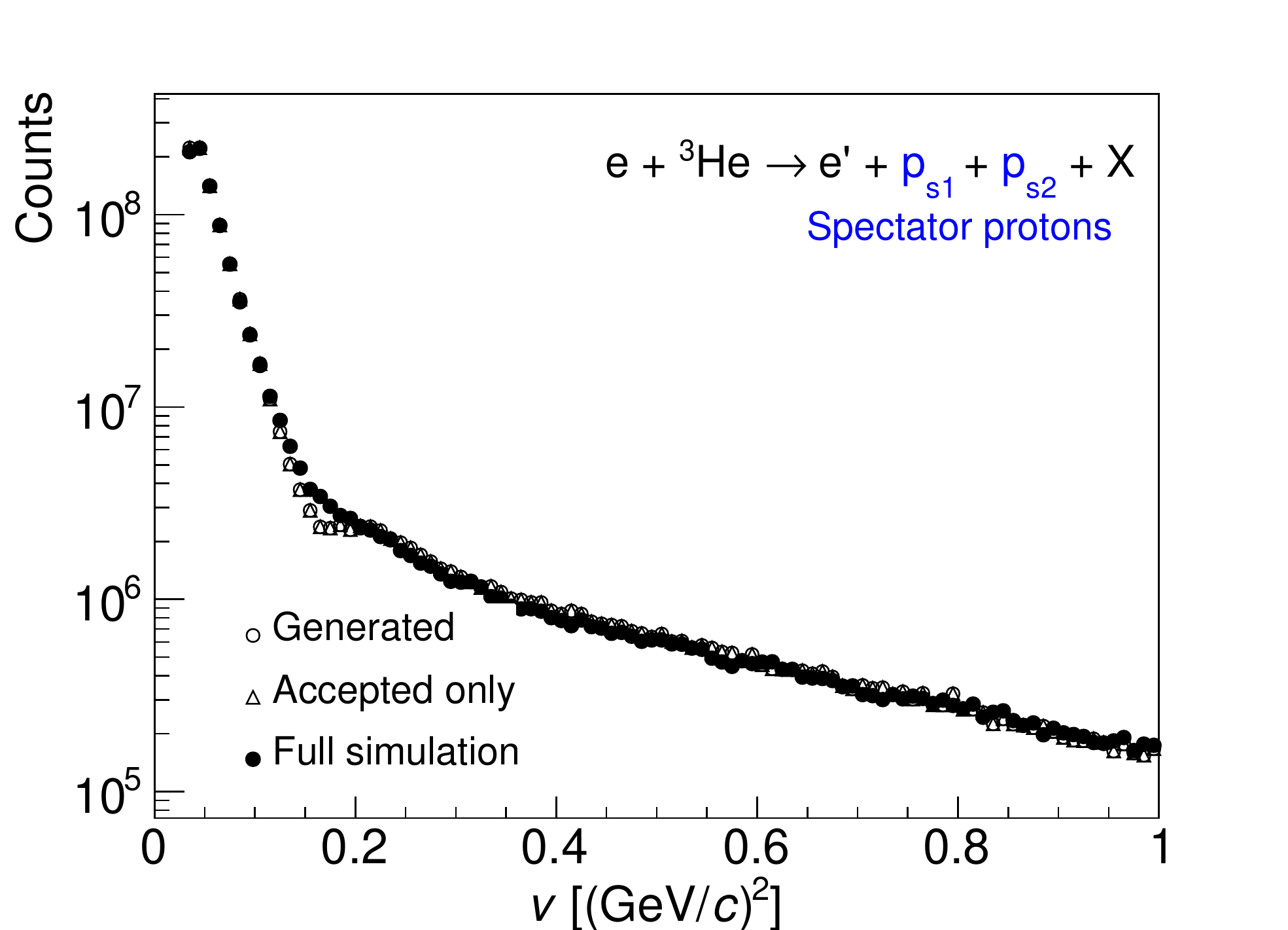}
    \caption{Distribution of the virtuality ($\nu$) for the kinematic setting 5x41. The simulation at the generated level, with acceptance correction and for the full simulation are shown as open circles, open triangles and the solid points, respectively}
    \label{vir-5x41}

%\end{figure}
%
%\begin{figure}[ht]
%    \centering
%    \includegraphics[width=\linewidth]{Figures/EIC_root_plots/vir_18x110.pdf}
%    \caption{Distribution of the virtuality ($\nu$) for the kinematic setting 18x110. The simulation at the generated level, with acceptance correction and for the full simulation are shown as open circles, open triangles and the solid points, respectively}
%    \label{vir-18x110}
\end{figure}

\begin{figure*}[ht]
    \centering
    \includegraphics[width = 0.49\linewidth]{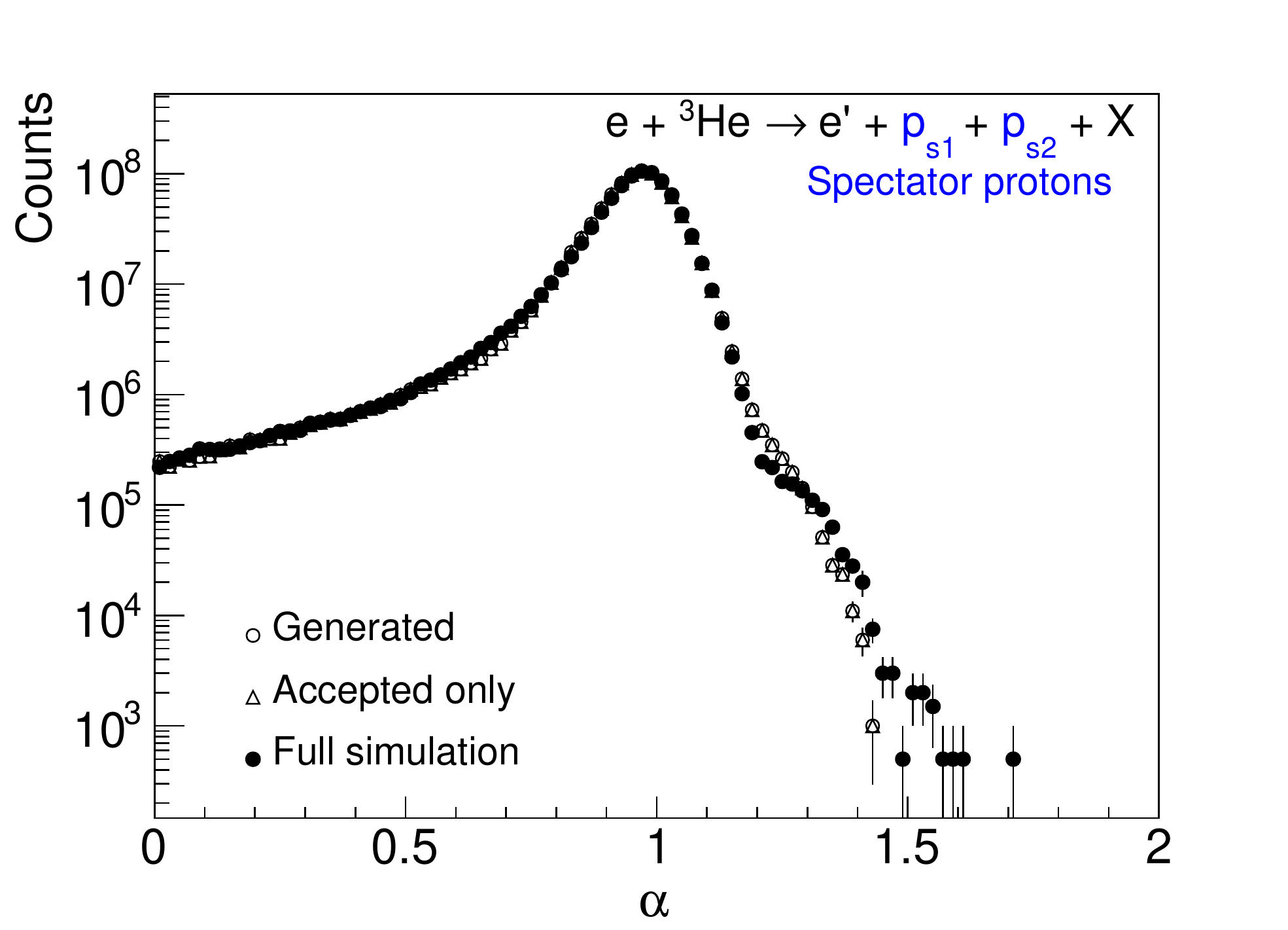}
    \includegraphics[width = 0.49\linewidth]{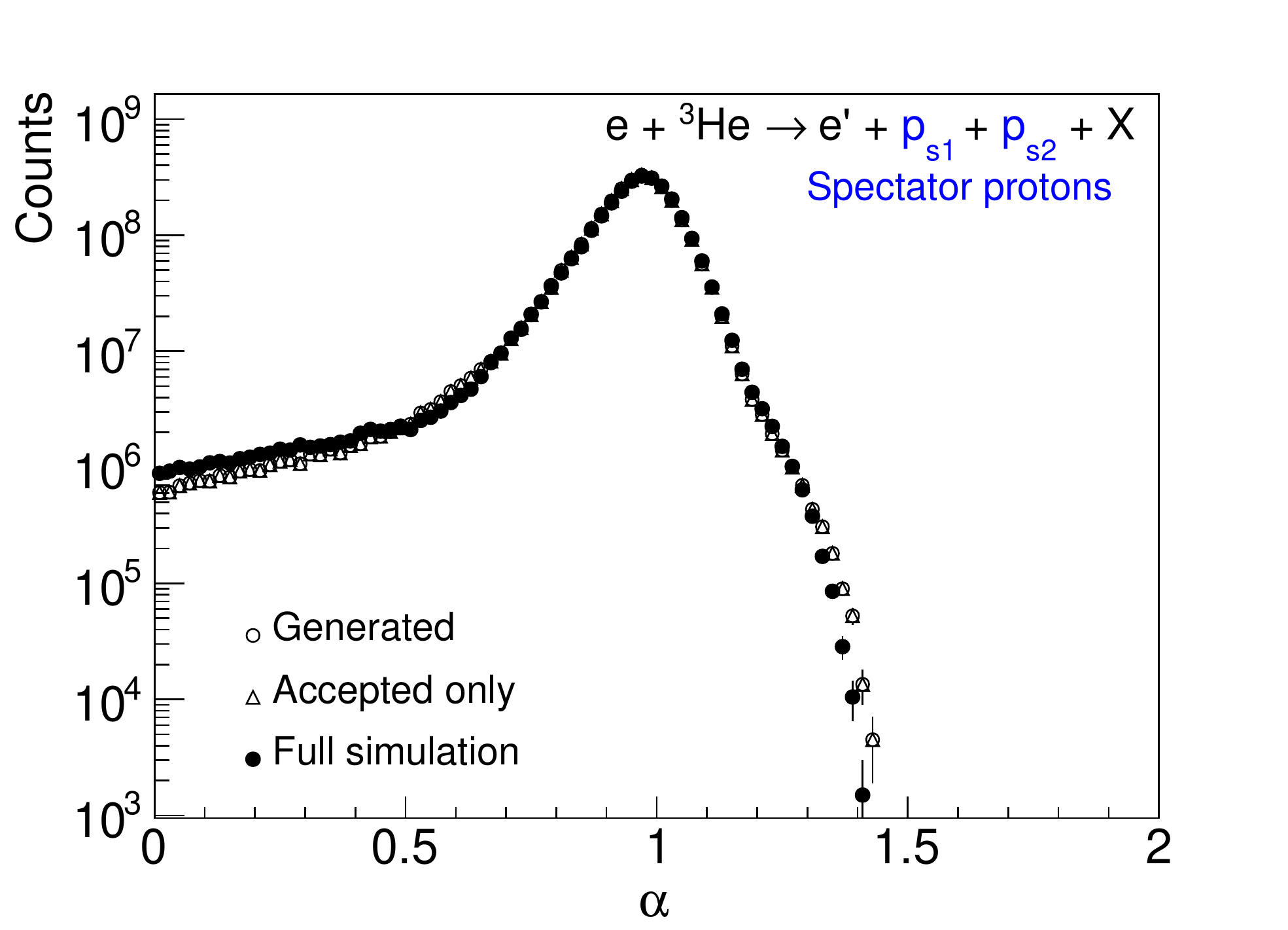}
    \caption{Distribution of the light-cone fraction ($alpha$) for the kinematic setting 5x41 GeV on the left and 18x110 GeV on the right. The simulation at the generated level, with acceptance correction and for the full simulation are shown as open circles, open triangles and the solid points, respectively.}
    \label{alpha-5x41}
\end{figure*}

\begin{figure}[ht]
    \centering
    \includegraphics[width = \linewidth]{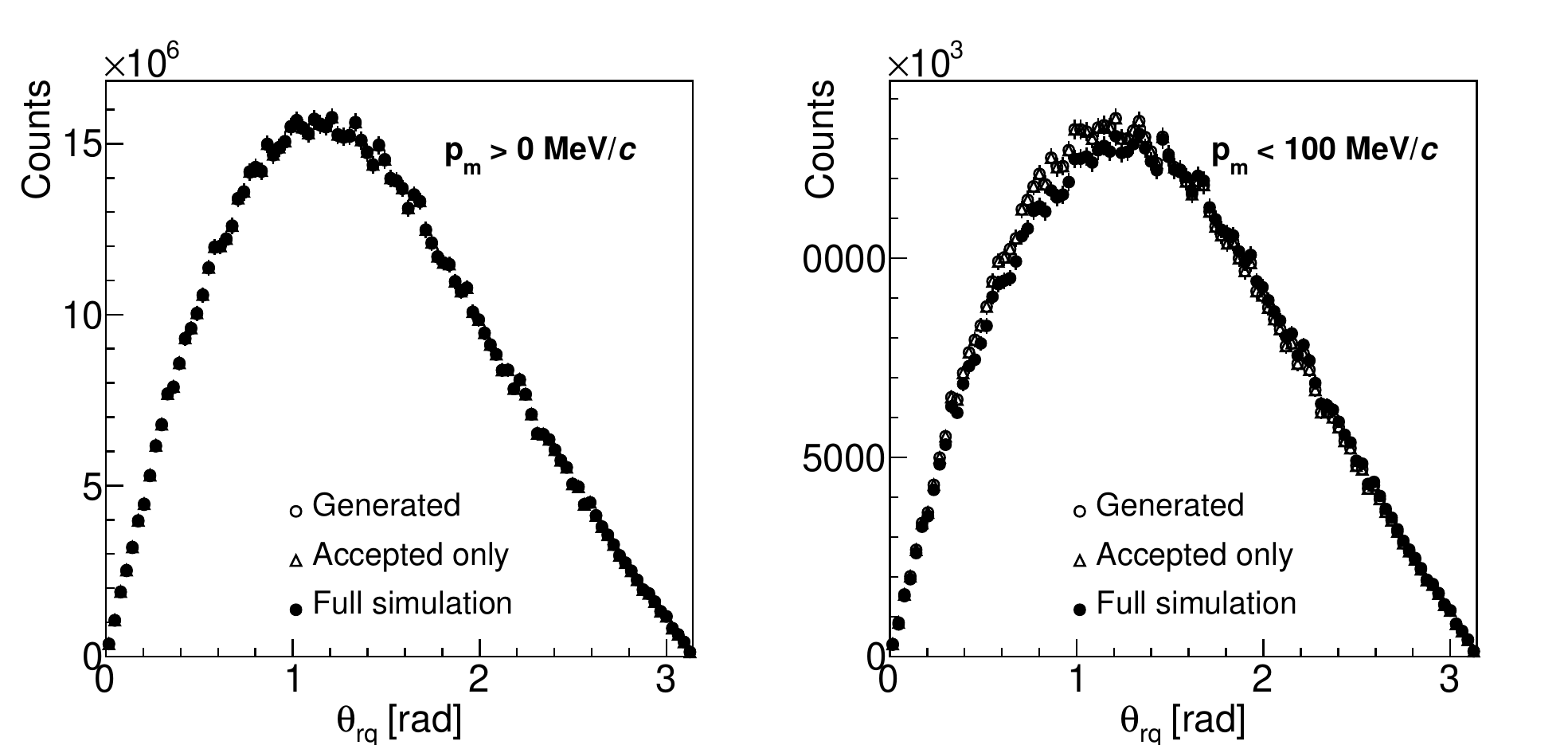}
    \caption{Distribution of the angle between the missing momentum vector (the sum of two spectator proton vectors) and the transfer momentum vector, for kinematic setting 5x41 GeV on the left and 18x110 GeV on the right in the IRF. The simulation at the generated level, with acceptance correction and for the full simulation are shown as open circles, open triangles and the solid points, respectively. }
    \label{theta-5x41}
\end{figure}

\begin{figure}[ht]
    \centering
    \includegraphics[width = \linewidth]{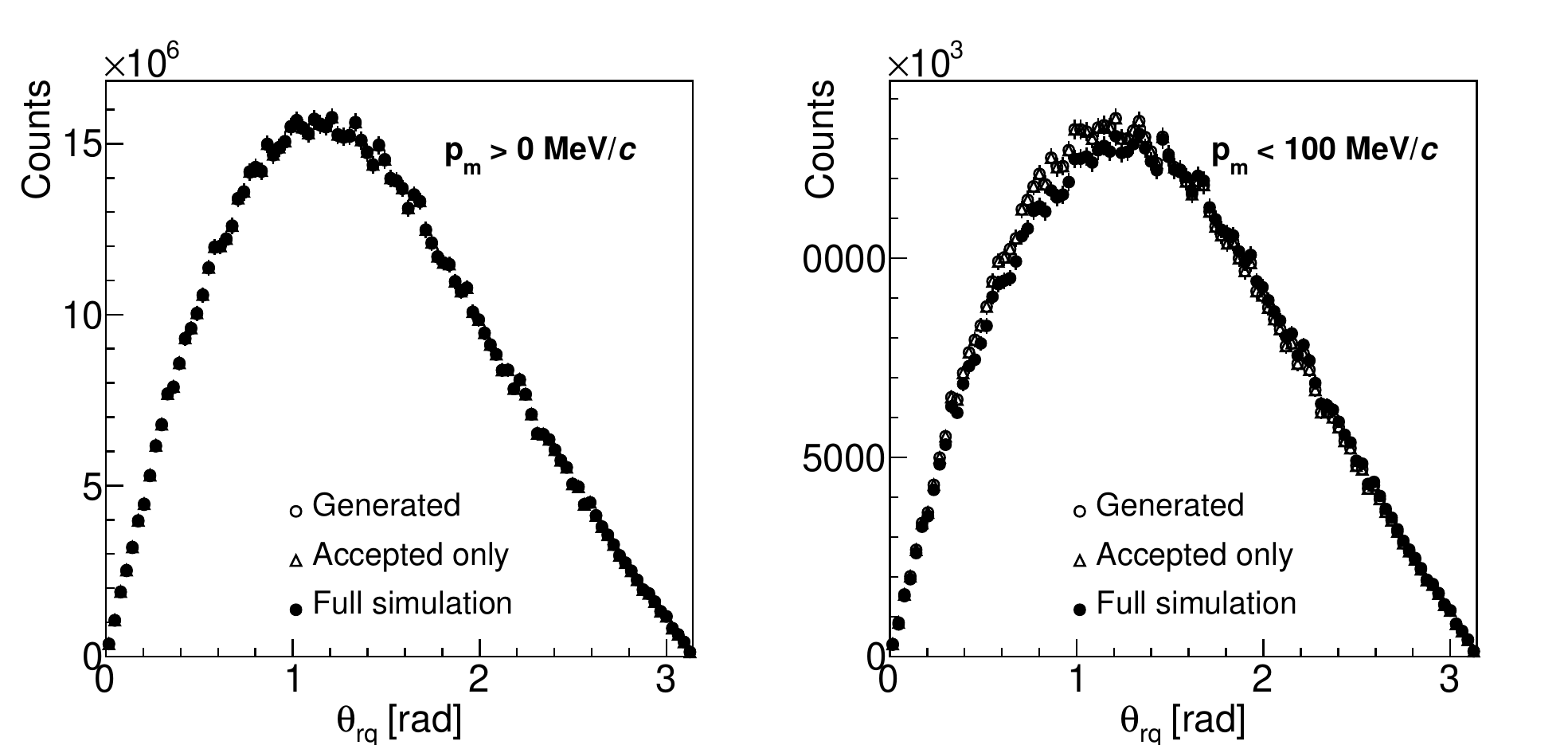}
    \caption{Distribution of the angle between the missing momentum vector (the sum of two spectator proton vectors) and the transfer momentum vector, for kinematic setting 18x110 in the IRF. The simulation at the generated level, with acceptance correction and for the full simulation are shown as open circles, open triangles and the solid points, respectively.}
    \label{theta-18x110}
\end{figure}

\section{Structure function $F_{2}^p$ and $F_2^D$ from NMC E155 fit}
\section{Parameters for $A_1^n$ and $A_1^p$ from the fitting}
\subsection{E155 fit}
The E155 fit gives parameter for $g_1^p/F_1^p$ and $g_1^n/F_1^n$ as following:

\begin{equation}
    \frac{g_1^p}{F_1^p} = x^{0.700}(0.817 + 1.014x - 1.489x^2)(1 - \frac{0.04}{Q^2})
\end{equation}

\begin{equation}
\frac{g_1^n}{F_1^n} = x^{-0.335}(-0.013 - 0.33x + 0.761x^2)(1 + \frac{0.13}{Q^2})
\end{equation}